\documentclass[showpacs,floatfix,twocolumn]{revtex4-1}
\usepackage{graphicx,psfrag,amsmath,amssymb,amsfonts,latexsym,color,epsf,dcolumn,graphpap}
\usepackage{caption}
\usepackage{subfig}
\usepackage{float}
\usepackage{lipsum}

\definecolor{red}{rgb}{1,0,0}
\definecolor{blue}{rgb}{0,0,1}
\definecolor{skyblue}{rgb}{0,0,.5}
\definecolor{green}{rgb}{0,1,0}
\definecolor{orange}{cmyk}{0,.4,1,0}

\newcommand{\fq}{\left(\frac{\hbar}{2 e}\right)^2}
\newcommand{\ELcav}{E_{L,\text{cav}}}

\begin{document}
\title{Dynamical Casimir effect in a double tunable superconducting circuit}
\author{F.C.~Lombardo$^{a}$, F.D.~Mazzitelli$^{b}$, A. Soba $^c$, P.I.~Villar $^a$}
\affiliation{
$^a$Departamento de F\'\i sica {\it Juan Jos\'e
Giambiagi}, FCEyN UBA and IFIBA CONICET-UBA, Facultad de Ciencias Exactas y Naturales,
Ciudad Universitaria, Pabell\' on I, 1428 Buenos Aires, Argentina.\\
$^b$Centro At\'omico Bariloche and Instituto Balseiro, 
Comisi\'on Nacional de Energ\'\i a At\'omica, R8402AGP Bariloche, Argentina.\\
$^c$Centro At\'omico Constituyentes, Comisi\'on Nacional de Energ\'\i a At\'omica, 
 Avenida General Paz 1499, San Mart\'\i n, Argentina}
\date{today}

\begin{abstract}
\noindent 
We present an analytical and numerical  analysis of the particle creation in a cavity ended with two SQUIDs, both subjected to time dependent magnetic fields. In the linear and lossless regime, the problem can be modeled  by a free quantum field in $1+1$ dimensions, in the presence  of boundary conditions that involve a time dependent linear combination of the field and its spatial and time derivatives. We consider a situation in which the boundary conditions at both ends are periodic functions of time, focusing on interesting features as the dependence of the rate of particle creation with the characteristics of the spectrum of the cavity, the conditions needed for parametric resonance, and interference phenomena due to simultaneous time dependence of the boundary conditions. We point out several concrete effects that could be tested experimentally.

\end{abstract}
\maketitle
\section{Introduction}\label{sec:intro}

In the presence of time dependent environments,  a quantum field initially in its vacuum state evolves into an excited state containing  real particles. 
Particle creation due to time dependent external conditions can be achieved in very different setups, and is broadly named ``dynamical Casimir effect" (DCE)\cite{Moore, Dodonov2010, Dalvit2011, Nation2012}.

The initial literature on this subject was focused  in the study of particle creation in the presence of ``moving mirrors", which impose boundary conditions at their position. The acceleration of the mirror induces nontrivial modifications to the normal modes of the electromagnetic field, and create photons from an initial vacuum state. 
However, the experimental verification of this effect is rather difficult, because the rate of particle production is in  general extremely small.
There have been several alternative proposals more appropriate to measure the DCE in the broad sense mentioned above, i.e. photon production
in the presence of time dependent environments \cite{Yablo,Braggio,Belgiorno,analogue,Nori}. 

Some years ago, the DCE has been experimentally observed in a superconducting waveguide ended with a 
superconducting quantum interference device (SQUID) \cite{Wilson}. The time dependent external conditions 
are produced by applying a time dependent magnetic flux through the SQUID. This generates a time dependent inductance, 
which in turn produces a time dependent boundary condition for the field in the waveguide \cite{Nori}. Under certain 
conditions,  this setup mimics that of an electromagnetic field in a waveguide ended with a moving mirror (see however \cite{Louko2015}).
The DCE has also been measured using an array of SQUIDs, that simulates a time dependent refraction index, see Ref.\cite{Paraoanu}. In Ref. \cite{farina14}, authors found, in the context of superconducting circuits, that appropriate adjustments of the parameters used in the SQUID experiment reveal remarkable predictions as unexpected nonparabolic spectral distributions and enhancement in the created particles. 

A simple variant of the proposal of Ref.\cite{Nori} is to consider
a superconducting cavity of finite size that is, a waveguide ended with two SQUIDs. In the static situation, when the SQUIDs are subjected to constant magnetic fluxes, the boundary conditions on two points at a finite distance on the waveguide produce a discrete spectrum. Therefore, when turning on
time dependent boundary conditions,  it is possible to tune the external frequency in order to have parametric amplification, in the same fashion as 
for finite size cavities with moving mirrors.  However, the boundary conditions for the field in the superconducting cavity ended with SQUIDs are qualitatively different from those of the electromagnetic field in the presence of mirrors, since they  may involve second time derivatives of the field. 
Therefore, both the static spectrum and the rate of particle creation  have a richer structure in this case. 

In a previous paper \cite{1squid}, we presented an analysis of this problem, in the particular case in which only one of the boundary conditions is time dependent. We have shown that,  after introducing appropriate boundary conditions, the field in the cavity can be described by a system of coupled harmonic oscillators, with time-dependent frequencies and couplings. We obtained the spectrum of the stationary cavity in terms of the parameters that define these boundary conditions,  and computed numerically the particle creation rates, with emphasis in their dependence with the properties of the static spectrum. 

In a recent work, Svensson et al \cite{suecos17}  initiated the experimental study of a double tunable cavity, in which both  ends are subjected to  time dependent boundary conditions. It has been shown experimentally that the double cavity shows some of the features predicted for the case of two moving mirrors \cite{Dalvit2}, particularly the fact that there could be destructive or constructive  interference depending on the relative phase of the excitations at both ends of the cavity. Other non-ideal aspects of the experimental results are less clear and deserve further analysis.

The theoretical aspects of the particle creation by two moving mirrors has been originally analyzed in the context of $1+1$ quantum fields satisfying Dirichlet boundary conditions at their positions \cite{Dalvit2}. In that case, the cavity has an equidistant spectrum, and some features of the DCE
are very different from that of a cavity with non-equidistant spectrum,  due to the fact that all modes become  coupled at resonant frequencies.
A numerical analysis for Dirichlet mirrors in $1+1$ and $3+1$ dimensions has been reported in Ref.\cite{2walls}, where it was shown that
the interference effects are also present for non-equidistant spectra.
However, as already mentioned, the waveguide with time dependent boundary conditions
has qualitative differences with respect to the cavity with moving mirrors.
It is then worth to analyze in detail the specific case of the double tunable cavity from a theoretical point of view. This is the aim of the present work. 

The paper is organized as follows.
 In Section II we describe the model for a (linearized) superconducting cavity  with time dependent boundary conditions, and show that the system can be described as a set of coupled harmonic oscillators. In Section III we study analytically  the particle creation rate using multiple scale analysis (MSA). 
We pay particular attention to the dependence of the results
with the main characteristics of the spectrum, to the existence of constructive and destructive interference,  and to the conditions under which the system enters a regime of parametric resonance. Section IV contains a numerical analysis of the spectrum of the static cavity. As we will see,
with appropriate choices of the parameters of the SQUIDs,  it is possible to generate  equidistant or 
non-equidistant spectra.  Section V is devoted to the numerical calculation of the particle creation rates. In addition to provide 
support to the analytic calculations of Section III, we explore regimes which are non reachable with the lowest order MSA (like
oscillations with large amplitudes, and the very long time behavior) and regimes that,  although in principle treatable with
MSA at higher orders (like the non leading resonances), are rather cumbersome to implement analytically. In Section VI we  study
the dependence of the results with the detuning of the external frequencies,  an important
aspect for the eventual experimental verification of these effects. Section VII contains the conclusions of our work.

\section{Doubly tunable superconducting cavity}


We shall consider a superconducting tunable resonator of length $d$, with a SQUID in each end, i.e.  at $x=0$ and at $x=d$. The idea is to have  
two independently controllable boundary conditions. 

For the theoretical description we 
extend previous results in Ref.\cite{1squid,Shumeiko}.
The cavity, which is assumed to have same capacitances $C_0$ and inductances $L_0$ per unit length,  for the 
both SQUIDs located at $x = 0$ and $x= d$ respectively, is described by the superconducting phase field $\phi(x,t)$ with Lagrangian
\begin{eqnarray}\label{lag}
L_{\rm cav} &=& \fq \frac{C_0}{2} \int_0^d d x \left(\dot \phi^2 - v^2 \phi'^2 \right) \nonumber \\
&+& \left[ \fq \frac{2 C^L_J}{2} \dot \phi_0^2 
 -  E^L_J \cos{f^L(t)} \phi_0^2
 \right] \nonumber \\
&+&\left[ \fq \frac{2 C^R_J}{2} \dot \phi_d^2 
 -  E^R_J \cos{f^R(t)} \phi_d^2
 \right] 
 \,,
\end{eqnarray}
where  $L$ and $R$ denote the SQUID in the left $x = 0$ boundary and the one on the right at $x = d$. In Eq.(\ref{lag}) we have set 
$v = 1/\sqrt{L_0 C_0}$ as the field propagation velocity,  and $\phi_{0,d}$ as the value of the field at the boundaries $\phi(0,t)$ and $\phi(d,t)$. $f^{L,R}(t)$ 
is the phase across the SQUIDs controlled by external magnetic fluxes. $E^{L,R}_J$ and $C^{L,R}_J$ denote the Josephson energies and capacitances, respectively (we will set $C^L_J = C^R_L = C_J)$. 
The Lagrangian in Eq.(\ref{lag}) contains additional contributions proportional to higher powers of $\phi_0$ and $\phi_d$ that  will not be considered in the rest of this paper. In what follows we will set $v=1$.

As anticipated, the description of the cavity involves the field $\phi(x,t)$ for $0<x<d$ and the  additional degree of freedom $\phi_{0,d}$. The dynamical equation reads
\begin{equation}
\ddot \phi - \phi'' = 0
\,,
\end{equation}
and the boundary conditions are

\begin{equation}\label{eqphi0}
\frac{\hbar^2}{E_C} \ddot \phi_0 + 2 E^L_J \cos{f^L(t)} \phi_0 + E_{\rm L,cav }d \phi'_0  = 0
\,,
\end{equation}
at $x = 0$ and

\begin{equation}\label{eqphid}
\frac{\hbar^2}{E_C} \ddot \phi_d + 2 E^R_J \cos{f^R(t)} \phi_d + E_{\rm L,cav} d \phi'_d  = 0
\,,
\end{equation}
at $x = d$. In these equations we have defined $E_C = (2e)^2/(2 C_J)$ and $\ELcav = (\hbar/2e)^2 (1/L_0 d)$. The equations 
above come from the variation of the action with respect to $\phi_{0,d}$,
and can be considered as a generalized boundary condition for the field.   The presence of second time derivatives of the field
pinpoints the existence of degrees of freedom localized on the boundary \cite{Fosco2013}.

As usual, it will be useful to write the Lagrangian in terms of eigenfunctions of the static cavity. Assuming that
\begin{equation}
f^{L,R}(t)=f^{L,R}_0 +  \theta (t) \theta(t_F-t)\epsilon_{L,R}\sin(\Omega_{L,R} t+\phi_{L,R})\, ,\label{pert}
\end{equation}
we can expand the field as
\begin{equation}\label{modeexpansion}
\phi(x,t) = {2e \over \hbar} \sqrt{2 \over C_0 d} \sum_n q_n(t) \cos\left( k_nx + \varphi_n\right)
\,,
\end{equation}
where the eigenfrequencies $k_n$ and the phases $\varphi_n$ satisfy Eqs. (\ref{eqphi0}) and (\ref{eqphid}) in the static case $f^{L,R}=f^{L,R}_0$:
\begin{eqnarray}
\label{spectrum}
k_n d\tan{(k_n d+\varphi_n)}  &=&   \frac{2 E^R_J \cos{f^R_0} }{E_{\rm L,cav}}  - \frac{2 C_J }{C_0 d}(k_n d)^2 \nonumber \\
k_n d\tan{\varphi_n}  &=& -    \frac{2 E^L_J \cos{f^L_0} }{E_{\rm L,cav}} + \frac{2 C_J }{C_0 d}(k_n d)^2. \label{tras}
\end{eqnarray}

Following previous developments for one-SQUID tunable cavity \cite{Shumeiko,1squid}, the dynamical 
equation for the mode $n$ is therefore written as

\begin{eqnarray}\label{eqmodes}
\ddot{q}_n &+& k_n^2 q_n = \frac{2 V^R_0}{d^2 M_n} \epsilon_R \theta(t)\theta(t_F-t) \sin (f^R_0) \nonumber \\
&\times & \sin(\Omega_R t+\phi_R) \cos(k_n d + \varphi_n) \sum_m q_m(t) \cos \left(k_md + \varphi_m\right) \nonumber \\
& + & \frac{2 V^L_0}{d^2M_n} \epsilon_L \theta(t)\theta(t_F-t) \sin (f^L_0) \nonumber \\
&\times & \sin(\Omega_L t+\phi_L) \cos \varphi_n  \sum_m q_m(t) \cos\varphi_m, 
\end{eqnarray}
where $V_0^{L,R} = 2 E_J^{L,R}/E_{\rm L,cav}$ and  we assumed that $\epsilon_{L,R}\ll 1$.  
We have also defined 
\begin{eqnarray}\label{Mn}
M_n &=& 1 + {\sin{\left[2\left(k_n d + \varphi_n\right)\right]} \over 2k_nd} - \frac{\sin 2\varphi_n}{2 k_n d} \nonumber \\
&+& 2 \chi_0 \cos^2\left(k_nd + \varphi_n\right)
\,,
\end{eqnarray}
where $\chi_0 = 2 C_J/(C_0 d)$.

The classical description of the theory consists of a set of coupled harmonic oscillators
with time dependent frequencies \cite{param1,param2}. The quantization of the system is straightforward. In the Heisenberg representation,
the variables $q_n(t)$ become quantum operators

\begin{equation}
\hat q_n(t)= \sum_{m} \frac{1}{\sqrt{2k_m}} \left[\hat a_m \epsilon_n^{(m)}(t) + \hat a^{\dagger}_m \epsilon_n^{(m)*}(t)\right],
\end{equation}
where $\hat a_m$ and  $\hat a^{\dagger}_m$ are the annihilation and creation operators. The functions $\epsilon_n^{(m)}(t)$ are 
properly normalized solutions of Eq.(\ref{eqmodes}), satisfying initial conditions 

\begin{eqnarray}
 \epsilon_n^{(m)}(t= 0) &=& \delta_{nm}, \nonumber \\
{\dot \epsilon}_n^{(m)} (t= 0) &=& -i k_n \delta_{nm}\, .
\nonumber \end{eqnarray} 

In the static regions $t<0$ and $t>t_F$ these functions are linear combinations of $e^{\pm i k_n t}$.  We define the {\it in}-basis as the 
solutions of Eq.(\ref{eqmodes}) that satisfy
\begin{equation}
\epsilon_n^{(m),in}(t)=e^{-i k_n t} \delta_{nm}\,\,\, \text {for}\,\,\,  t<0 \,\, .\label{eps0}
\end{equation}
The associated annihilation operators $a_n^{in}$ define the {\it in}-vacuum $\vert 0_{in}\rangle$. The {\it out }-basis 
$\epsilon_n^{(m),out}$ is introduced in a similar way, defining the behaviour  for $t>t_F$. 
The {\it in} and {\it out} basis are connected by a Bogoliubov transformation
\begin{equation}
\epsilon_n^{(m)}(t)=\alpha_{nm}  e^{-i k_n t}  +\beta_{nm} e^{i k_n t} \, ,\label{epst}
\end{equation}
and the number of created particles in the mode $n$ for $t>t_F$ is given by \cite{error}
\begin{equation}
N_n=\langle 0_{in}\vert a_n^{out\, \dagger} a_n^{out}\vert0_{in}\rangle =\sum_m \vert\beta_{nm}\vert^2\, . \label{Npart}
\end {equation}

In the present paper, we shall numerically solve the dynamical Eqs.(\ref{eqmodes}) and evaluate the number of created particles using 
Eq.(\ref{Npart}).  Before doing that, we will present an analytic study which is appropriate for resonant external frequencies.

\section{Analytic results: multiple scale analysis}

In order to study analytically  Eqs.(\ref{eqmodes}) we write them in the form
\begin{eqnarray}\label{eqmodecomp}
\ddot q_n +\omega_n^2(t)q_n=\sum_{m\neq n} \sum_{j}q_m S^{(j)}_{nm}\sin(\Omega_j t+\phi_j)\, ,
\end{eqnarray}
where $j=L,R$, we made the redefinition $q_n\to q_n/\sqrt{M_n}$,  and
\begin{eqnarray}
\omega^2_n(t)&=&k^2_n -\sum_{j}\alpha^{(j)}_n \sin(\Omega_j t+\phi_j) \nonumber\\
S^R_{mn}&=& \frac{2V_0^R}{d^2\sqrt{M_nM_m}} \epsilon_R \sin f^R(0) \cos\left(k_n d + \varphi_n\right) \nonumber \\
&\times & \cos\left(k_m d + \varphi_m\right)\nonumber\\
S^L_{mn}&=& \frac{2V_0^L}{d^2\sqrt{M_nM_m}} \epsilon_L \sin f^L(0) \cos\varphi_n \cos\varphi_m\nonumber\\
\alpha_n^R &=& \frac{2V_0^{R}}{d^2M_n} \epsilon_{R} \sin f^R(0)\cos^2(k_n d + \varphi_n) \nonumber \\
\alpha_n^L &=& \frac{2V_0^{L}}{d^2M_n} \epsilon_{L} \sin f^L(0)\cos^2(\varphi_n) .
\label{ec16}
\end{eqnarray} 
We will assume that the amplitude of the 
time dependence is small, that is $\alpha\ll 1$.  We will also set $\epsilon_R=\epsilon_L=\epsilon$. 

It is known that, due to parametric resonance,  a naive perturbative solution of Eqs.(\ref{eqmodecomp})
in powers of $\epsilon$ breaks down after a short amount of time.
In order  to find a solution valid for longer times  we use the multiple scale analysis (MSA) technique \cite{bender, crocces}. 
We introduce a second timescale $\tau=\epsilon t$, and write
\begin{equation}
q_n(t,\tau)=A_n(\tau)\frac{e^{-i k_n t}}{\sqrt{2 k_n}}+B_n(\tau)\frac{e^{i k_n t}}{\sqrt{2 k_n}}\, . \label{ec17}
\end{equation}
The functions $A_n$ and $B_n$ are slowly varying, and contain the cumulative resonant effects. To obtain differential equations for them, we insert 
this ansatz into Eq.(\ref{eqmodecomp}) and neglect second derivatives of $A_n$ and $B_n$. After multiplying the equation by 
$\exp{(\pm i k_n t)}$, and averaging over the fast oscillations we obtain
\begin{widetext}
\begin{eqnarray}\label{eqsAyB}
4k_n \frac{dA_n}{dt}&=& -B_n \sum_j\alpha_n^{(j)}\delta(\Omega_j-2k_n)e^{-i\phi_j}
+ \sum_{m\neq n}\sum_j S_{mn}^{(j)}[A_m \left(\delta(\Omega_j-k_m+k_n)e^{i\phi_j} - \delta(\Omega_j+k_m-k_n)e^{-i\phi_j} \right) 
\nonumber\\
&& - B_m \delta(\Omega_j-k_n-k_m)  e^{-i\phi_j} ]\, ,\nonumber\\
4k_n \frac{dB_n}{dt}&=& -A_n \sum_j\alpha_n^{(j)}\delta(\Omega_j-2k_n)e^{i\phi_j}
- \sum_{m\neq n}\sum_j S_{mn}^{(j)}[B_m \left(\delta(\Omega_j+k_m-k_n)e^{i\phi_j} - \delta(\Omega_j+k_n-k_m)e^{-i\phi_j} \right) 
\nonumber\\
&& + A_m \delta(\Omega_j-k_n-k_m)  e^{i\phi_j} ]\, ,
\end{eqnarray}
\end{widetext}
where $\delta (x)$ should be understood as a Kronecker delta $\delta_{x 0}$.

In the above equations, the phase $\phi_j$ express the dephasing between the harmonic external excitations at right and left squids.
Assuming that $\sin f^R(0)$ and $\sin f^L(0)$ have the same sign,  if $\phi_R-\phi_L=0$ the SQUIDs are out of phase. We refer to this as the breathing mode. On the contrary, when $\phi_R-\phi_L= \pi$, we will find the so called shaker modes (electromagnetic shaker in the case of a cavity with two oscillating mirrors).

We can see that these equations are non trivial when the
 external harmonic driving frequencies are just tuned with one eigenvalue of the static cavity $\Omega_{L,R}= 2k_n$. Moreover, 
other modes will be coupled and will resonate if the conditions 
 \begin{equation}\label{rescondition}
 \Omega_{L,R} = \vert k_n\pm k_j\vert
 \end{equation}
 are satisfied. We will now describe some particular cases.
 
 \subsection{A single resonant mode}

 We assume that $\Omega_L=\Omega_R=2 k_n$ for some mode, and that no other resonant condition is satisfied. In this case
 the dynamical equations Eq.\eqref{eqsAyB} reduce to 
 \begin{eqnarray}\label{1mode}
 4k_n \frac{dA_n}{dt}&=& -B_n[\alpha_n^L+\alpha_n^Re^{-i\phi_R}]\nonumber\\
4k_n \frac{dB_n}{d t}&=& -A_n[\alpha_n^L+\alpha_n^Re^{-i\phi_R}] \, ,
\end{eqnarray}
where we have assumed that $\phi_L=0$.
From these equations, it is easy to see that the number of created particles grows exponentially with a rate $\Gamma_n$ given by 
\begin{equation}
\Gamma_n = \frac{1}{4 k_n} \sqrt{(\alpha_n^R)^2+(\alpha_n^L)^2+ 2 \alpha_n^R\alpha_n^L \cos\phi_R}\, .
\label{onemode}\end{equation}
As expected from calculations of the DCE for mirrors, there is constructive interference for $\phi_R =0$, and destructive for $\phi_R =\pi$. 
Note however that even in the case $\epsilon_R=\epsilon_L$ considered here, the interference is partial, due to the fact that in general
$\alpha_n^R\neq \alpha_n^L$. 

\subsection {Two resonant modes}

We now assume that the only resonant condition satisfied by the external frequency is $\Omega_L=\Omega_R=k_m+k_n$, for a couple of modes $n$ and $m$. The dynamical equations read, in this case, as
 \begin{eqnarray}\label{2modes}
 4k_n \frac{dA_n}{dt}&=& -B_m (S_{mn}^L+S_{mn}^R e^{-i\phi_R})\nonumber\\
 4k_m \frac{dB_m}{dt}&=& -A_n (S_{mn}^L+S_{mn}^R e^{i\phi_R})\nonumber\\
4k_m \frac{dA_m}{dt}&=& -B_n (S_{mn}^L+S_{mn}^R e^{-i\phi_R})\nonumber\\
4k_n \frac{dB_n}{dt}&=& -A_m (S_{mn}^L+S_{mn}^R e^{i\phi_R})\, .
\end{eqnarray}
Combining these equations it is easy to show
that all the functions grow exponentially with a rate 
\begin{equation}
\frac{\vert\Gamma_{mn}\vert}{4\sqrt{k_mk_n}}
\end{equation}
where
\begin{equation}
\Gamma_{mn}=S_{mn}^L+S_{mn}^R e^{-i\phi_R}\, .
\end{equation}
Therefore, the number of created particles grows exponentially in both modes, with a rate that depends on the dephasing of the harmonic 
external excitation.

It is interesting to remark that the case  $\Omega_L=\Omega_R=k_m - k_n$\
is qualitatively different (we assume $k_m>k_n$).  We have
 \begin{eqnarray}\label{2modes2}
 4k_n \frac{dA_n}{dt}&=& A_m (S_{mn}^L+S_{mn}^R e^{i\phi_R})\nonumber\\
 4k_m \frac{dA_m}{dt}&=& -A_n (S_{mn}^L+S_{mn}^R e^{-i\phi_R})\, ,
 \end{eqnarray}
 and 
  \begin{eqnarray}\label{2modes3}
 4k_n \frac{dB_n}{dt}&=& B_m (S_{mn}^L+S_{mn}^R e^{-i\phi_R})\nonumber\\
 4k_m \frac{dB_m}{dt}&=& -B_n (S_{mn}^L+S_{mn}^R e^{i\phi_R})\, .
\end{eqnarray}
Note that in this case  there is no mixing between the coefficients $A_n$ and $B_n$. Moreover, due 
to the relative sign in the rhs of the equations, they lead to an oscillatory behavior.

\subsection{Two different external frequencies}

We will now consider cases in which $\Omega_R\neq \Omega_L$, but both still satisfying some of the resonant conditions.
The simplest choice is to tune each frequency with a different mode, that is $\Omega_L = 2 k_n$ and $\Omega_R=2 k_m$, with $m\neq n$.
In this case, there is no mode mixing, and each one resonates independently of the other.

More interesting situations are $(\Omega_L,\Omega_R)=(2k_n,k_m-k_n)$ and  $(\Omega_L,\Omega_R)=(k_m + k_n,k_m-k_n)$. In both cases,
the dynamical equations reduce to a system of four coupled differential equations (note that in the previous examples the equations 
are coupled in pairs).

We first consider $(\Omega_L,\Omega_R)=(2k_n,k_m-k_n)$. The equations read
\begin{eqnarray}\label{2freq1}
 4k_n \frac{dA_n}{dt}&=& -B_n\alpha_n^L e^{-i\phi_L} + A_m S_{mn}^R e^{i\phi_R}\nonumber\\
4k_m \frac{dA_m}{dt}&=& -B_m \alpha_m^L e^{-i\phi_L} - A_n S_{mn}^R e^{-i\phi_R}\nonumber\\
4k_n \frac{dB_n}{dt}&=& -A_n \alpha_n^L e^{i\phi_L} + B_m S_{mn}^R e^{-i\phi_R}\nonumber\\
4k_m \frac{dB_m}{dt}&=& -A_m \alpha_m^L e^{i\phi_L} - B_n S_{mn}^R e^{i\phi_R}\, .
\end{eqnarray}
Note that, due to the particular choice of the external frequencies, the equations involve both  the parameters $\alpha_n^L$ and $S_{mn}^R$. 
The solutions to this system of differential equations are of the form $\exp[\lambda_a t]$ where $\lambda_a$ $(a=1,2,3,4)$ are the
eigenvalues of the  $4\times 4$ matrix $M$ defined by 
\begin{equation}
\frac{d}{dt}\begin{bmatrix} A_n\\  A_m\\ B_n\\ B_m\end{bmatrix} = M \begin{bmatrix} A_n\\  A_m\\ B_n\\ B_m\end{bmatrix} . 
\end{equation}
For the particular case $\phi_L=0$, these eigenvalues are of the form
\begin{equation}
\lambda_a=\pm\sqrt{X\pm\sqrt{X_1+X_2\cos\phi_R}}\, ,
\end{equation}
where
\begin{eqnarray}
X &=& \left(\frac{\alpha_n^L}{4 k_n}\right)^2 + \left(\frac{\alpha_n^L}{4 k_m}\right)^2-\frac{(S_{mn}^R)^2}{16 k_mk_n} \nonumber\\
X_1 &=& \left[ \left(\frac{\alpha_n^L}{4 k_n}\right)^2-\left(\frac{\alpha_m^L}{4 k_m}\right)^2\right]^2 - \frac{(S_{mn}^R)^2}{4 k_mk_n}\times\nonumber\\
&&\left[ \left(\frac{\alpha_n^L}{4 k_n}\right)^2+\left(\frac{\alpha_m^L}{4 k_m}\right)^2\right]\nonumber\\
X_2 &=& -\frac{\alpha_n^L\alpha_m^L(S_{mn}^R)^2}{32 k_n^2k_m^2}\, .
\end{eqnarray}
Whether there is an eigenvalue with positive real part or not depends on the particular pair of frequencies considered, which determines 
the full set of parameters that define $X$, $X_1$ and $X_2$. The only case in which there are no resonant effects is when
the parameters are such that $X\pm\sqrt{X_1+X_2\cos\phi_R}<0$.

The case $(\Omega_L,\Omega_R)=(k_n+k_m,k_m-k_n)$ can be considered along similar lines. In this situation, the matrix $M$ depends on the coefficients
$S_{mn}^L$ and $S_{mn}^R$. As it is not possible to find analytic expressions for the eigenvalues, we omit the details. We will present a numerical example 
of this case in Section \ref{difOmegas}, showing that parametric amplification can occur.


\section{The  cavity spectrum}

Given the strong dependence of the particle creation rate with the spectrum of the static cavity, as can be seen from the analysis of the previous Section, it is important to analyze the spectra that
result from the generalized boundary conditions in the tunable superconducting cavity (Eq.(\ref{tras})). The 
spectrum is determined by the solution of that system of equations, that can be rewritten in terms of the new parameters of the 
cavity $\chi_0$ and $b_{0L,R}$ as 
\begin{eqnarray}
\label{spectrum2}
(k_n d)\tan{(k_n d + \varphi_n)}  + \chi_0 (k_n d)^2 &=&  b_{0R}  \nonumber \\
-(k_n d)\tan{\varphi_n} + \chi_0  (k_n d)^2  &=&  b_{0L} ,
\end{eqnarray}
where we have set $b_{0L,R} = V_0^{L,R} \cos f^{L,R}_0$.
The three free parameters that determine the solutions of Eq.(\ref{spectrum2}) are $\chi_0$, $b_{0R,L}$. 

Before describing the numerical study of these equations, let us discuss some general properties. Although not completely 
evident from Eq.\eqref{spectrum2}, the spectrum is symmetric under the interchange $b_{0L}\leftrightarrow b_{0R}$.
Indeed, one can show that if $(k_n,\varphi_n)$ solves Eq.\eqref{spectrum2}, then $(k_n,  -\varphi_n-k_nd)$ solves
the equations with $L\leftrightarrow R$. The spectrum does not change (the phases do).  

An important property that influence the rate of particle creation is whether the spectrum is equidistant or not. It is easy to see that 
for large values of both $b_0^L$ and $b_0^R$ and not so large values of $\chi_0$ and $k_n d$, the solutions  of Eq.\eqref{spectrum2}
are $k_n d \approx n\pi$. This is because for $b_{0L,R}\gg 1$ both  $\tan(k_n d + \varphi_n)$ and $\tan\varphi_n$ 
should be large numbers.

It is also easy to find situations where the spectrum is non-equidistant. For example, if $b_{0R} = b_{0L} = b_0$ we have
\begin{eqnarray}
&& k_n d \tan\left(\frac{n\pi - k_nd}{2}\right) - \chi_0 (k_nd)^2 = - b_0  \nonumber \\
&& k_n d + 2 \varphi_n = n \pi, \label{equalb0}
\end{eqnarray} without  an equidistant solution unless $b_0\gg1$.

\begin{figure}[h!]
\begin{center}
\includegraphics[width=8cm]{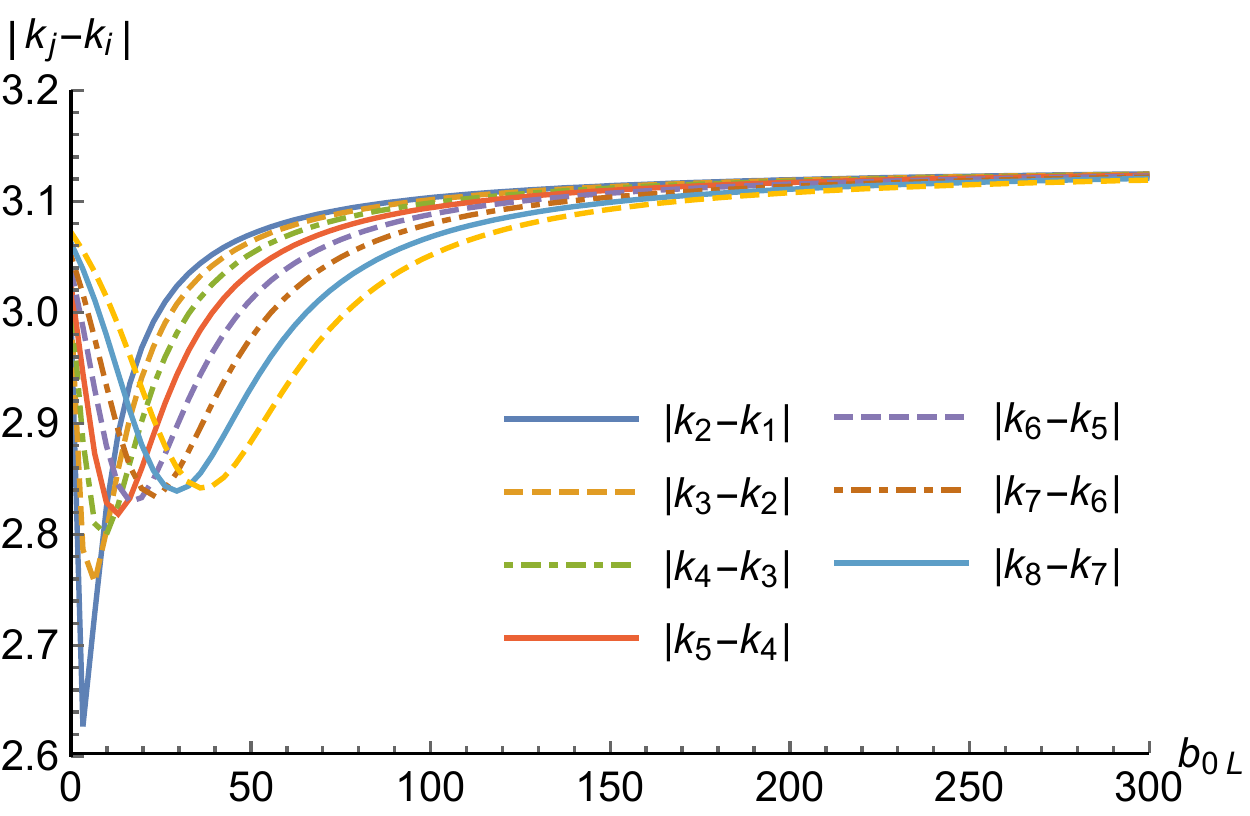}
\caption{(color online). Difference of consecutive eigenfrequencies  $|k_{j+1}-k_j|$ as a function of $b_{0L}$ for a fixed value of $\chi_0=0.05$ 
and $b_{0R}=500$ for the first twelve eigenfrequencies.}
\label{fig1}
\end{center}
\end{figure}

In order to obtain numerically the eigenfrequencies of the cavity from Eq.(\ref{spectrum2}) we use a single Newton-Raphson method with an stopping error of $10^{-6}$. 
In the first place, we shall study the difference between consecutive eigenfrequencies as a function of $b_{0L}$  for a typical experimental value 
\cite{Shumeiko}, say $\chi_0 = 0.05$ and fixed $b_{0R}$ ($b_{0R} = 500$). We can see in Fig.\ref{fig1} that the bigger the value of $b_{0L}$, the more equidistant is the spectrum for small consecutive eigenfrequencies. The difference between any consecutive eigenvalues of the cavity goes to a constant value of the order of $\pi$ when $b_{0L} \ge b_{0R}$.

\begin{figure}[h!]
\begin{center}
\includegraphics[width=8cm]{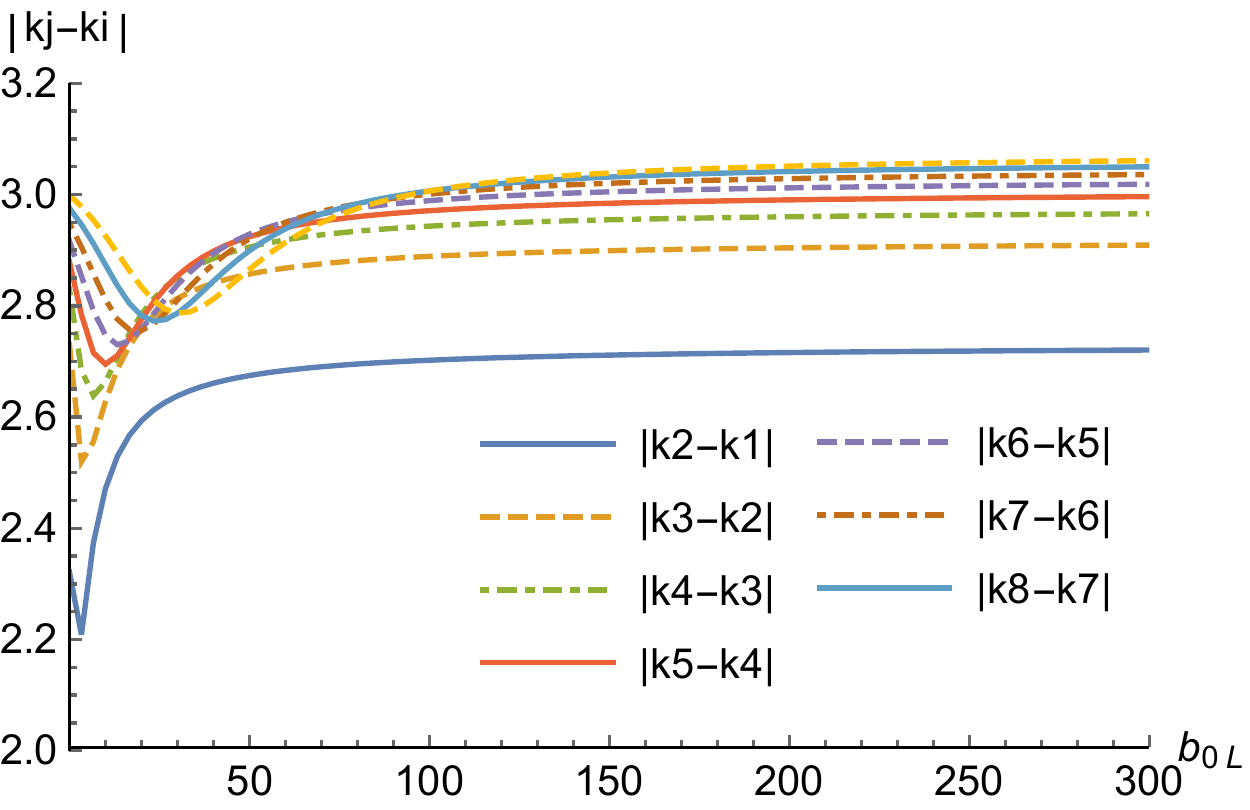}
\caption{(color online). Difference of consecutive eigenfrequencies $|k_{j+1}-k_j|$ as a function of $b_{0L}$ for a fixed value of $\chi_0=0.05$ and $b_{0R}=1$ in the case of a non-equidistant situation. We show the eigenfrequencies from $k_1$ up to $k_{10}$.}
\label{fig2}
\end{center}
\end{figure}

In Fig.\ref{fig2} we also show the difference of consecutive eigenfrequencies as a function of $b_{0L}$ for a smaller value of $b_{0R}=1$.

\begin{figure}
\begin{center}
                \includegraphics[width=8cm]{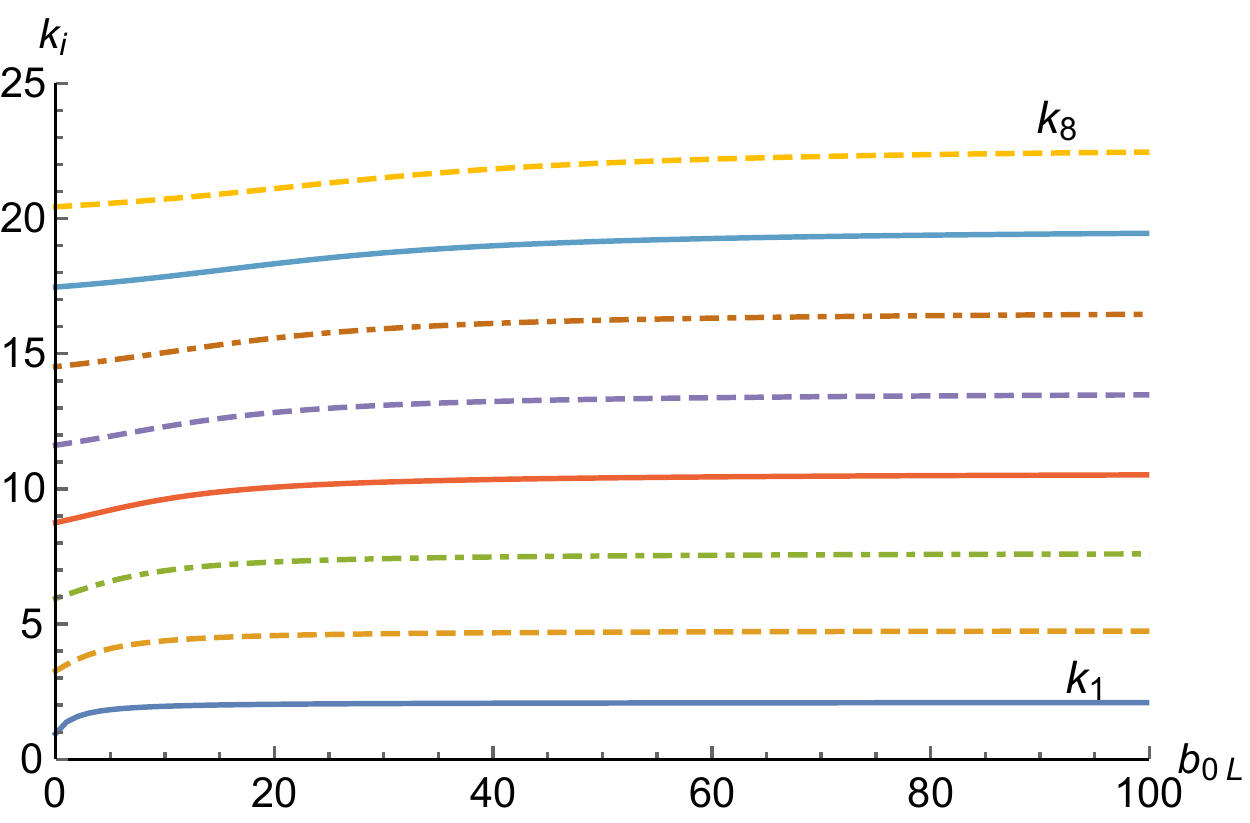}
                \caption{(color online). Eigenfrequencies $k_i$ as a function of $b_{0L}$, obtained for $b_{0R}=1$ with a fixed value of $\chi_0=0.05$}
                \label{fig3}
 \end{center}
                \end{figure}
        
           \begin{figure}
  \begin{center}
                \includegraphics[width=8cm]{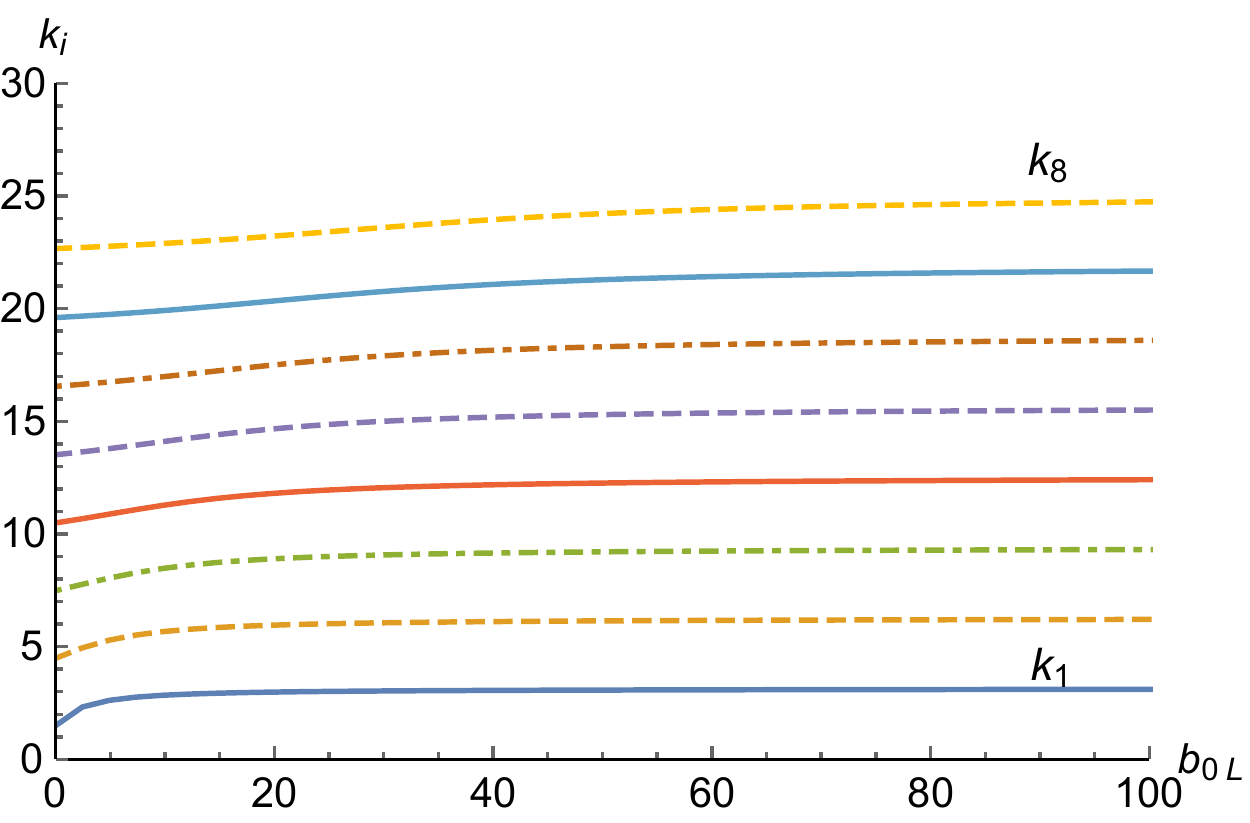}
                \caption{(color online). Eigenfrequencies $k_i$ as a function of $b_{0L}$, obtained for $b_{0R}=500$ with a fixed value of $\chi_0=0.05$ }
                \label{fig4}
  \end{center}
\end{figure}

\begin{figure}      
  \begin{center}
                \includegraphics[width=8cm]{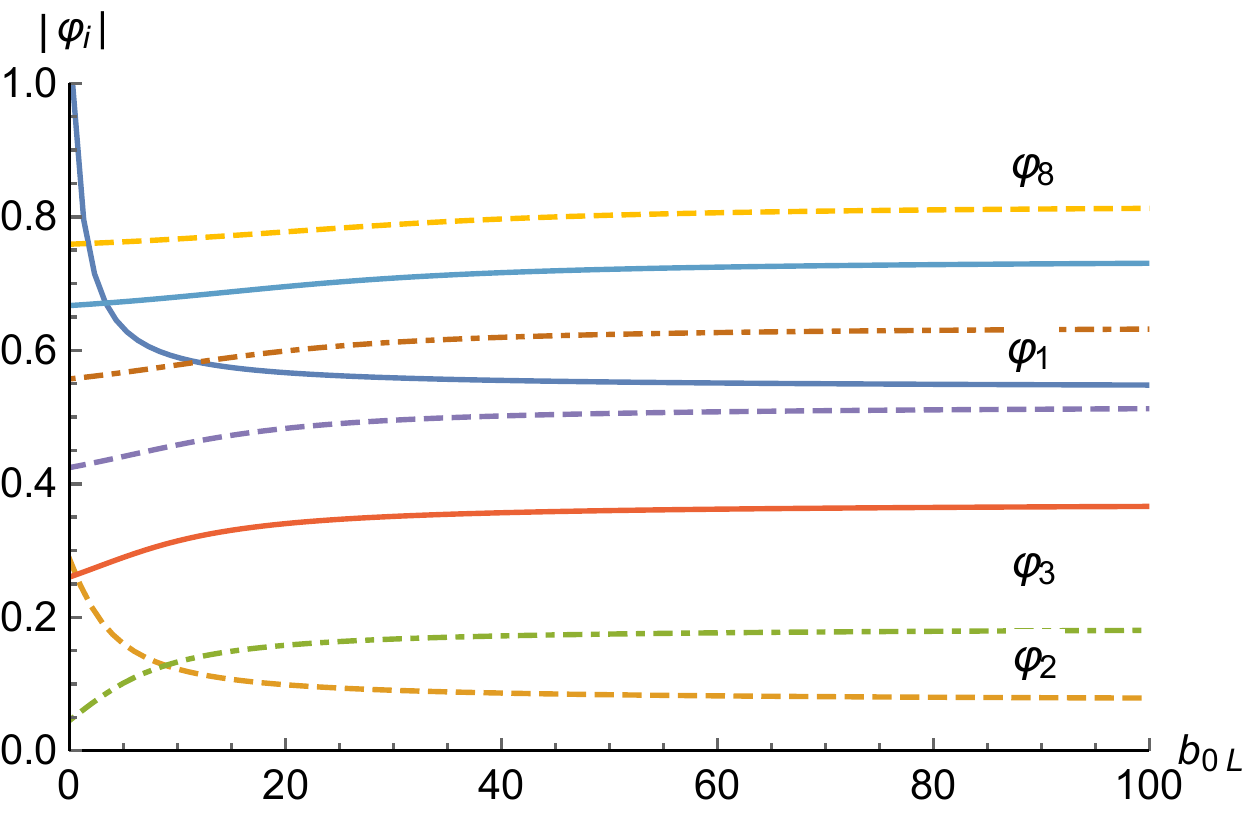}
                \caption{(color online). Phases $\varphi_n$ as a function of $b_{0L}$, obtained for $b_{0R}=1$ and $\chi_0=0.05$}
                \label{fig5}
       \end{center}
       \end{figure}

\begin{figure}
          \begin{center}
                \includegraphics[width=8cm]{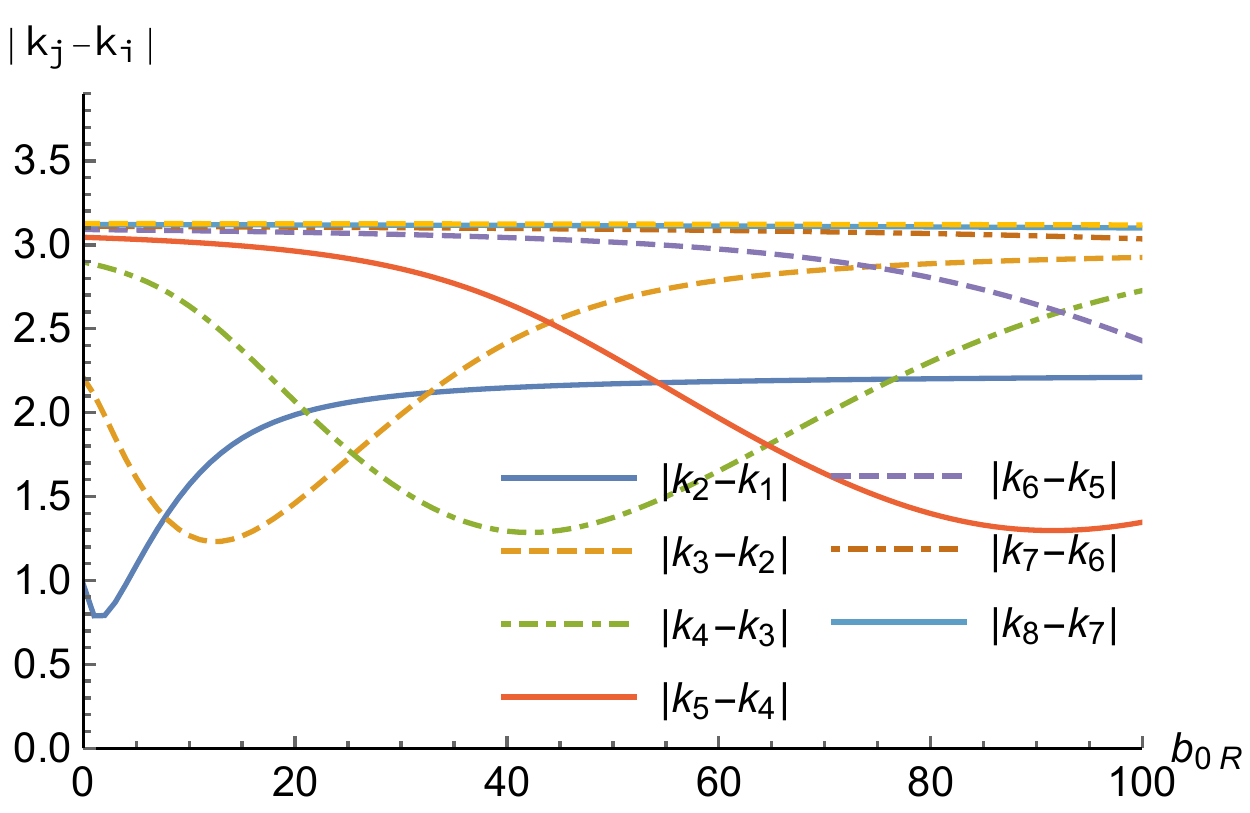}
               \caption{(color online). Difference $|k_{j}-k_i|$ between consecutive eigenfrequencies as a function of  $b_{0R}$ for a fixed value of $\chi_0=1$ and $b_{0L}=1$. The spectrum 
               presents major differences among the frequency even for higher order of frequencies at all values of $b_{0L,R}$}
                \label{fig6}
\end{center}
\end{figure}

In Fig.\ref{fig3}  we show the values of different consecutive eigenfrequencies as functions of  $b_{0L}$, for a fixed value of $b_{0R}=1$, while in Fig.\ref{fig4} we present the same eigenfrequencies when $b_{0R}=500$ and $\chi_0=0.05$ in both cases. In Fig.\ref{fig5}, we present the values of the phases obtained by solving Eq.(\ref{spectrum2})  for the case of $b_{0R}=1$ and $\chi_0=0.05$. In Fig. \ref{fig6}, we show the difference of consecutive eigenfrequencies $|k_{j+1}-k_j|$ as a function of $b_{0R}$ for a fixed value of $\chi_0=1$ and $b_{0L}=1$.

In summary, the results of this Section show that the spectrum of the doubly tunable cavity can be adjusted modifying the external (static) magnetic fluxes on the SQUIDs.

\section{Photon generation}

In this Section, we shall analyze different cases of photon generation by choosing particular values of the several parameters involved in the system's configuration. 
As described previously, the static spectrum is determined 
by the values of $\chi_0$, $b_{0L}$ and $b_{0R}$. The mode equations depend 
in addition on the static fluxes $f_0^R$ and $f_0^L$. The external excitation is described by the amplitude $\epsilon$, the frequencies $\Omega_R$ and $\Omega_L$ and the phases $\phi_R$ and $\phi_L$.

 It is appropriate to emphasize here that, in order not to vary so many parameters
of the static cavity, we are only going to vary  $b_{0L}$ and $b_{0R}$,  which
give enough freedom to produce qualitative changes in the spectrum. The other static parameters  are set to the particular values 
$\chi_0=0.05$, $f_0^R=f_0^L= 0.45 \pi$  unless explicitly indicated otherwise in the text.

\subsection{The numerical method}

In terms of the functions $\epsilon_n^{(m)}$ (that we will call $\epsilon_{nm}$ from now on), the equation of motion in Eq.(\ref{eqmodecomp}) can be written as
\begin{equation}
\ddot \epsilon_{nm} +\omega_n^2(t) \epsilon_{nm} = \sum_{j\neq n}\sigma_{nj}(t) \epsilon_{jm}\, ,\label{modeseps}
\end{equation}
or, equivalently
\begin{eqnarray}
\dot{\epsilon}_{nm} &=& U_{nm}, \nonumber \\
\dot{U}_{nm} &=& -\omega_n^2(t) \epsilon_{nm} - \sum_{j\neq n} \sigma_{n j} (t) \epsilon_{jm} , 
\end{eqnarray} where the explicit form of  $\sigma_{nj}(t)$ can be obtained  from Eqs. \eqref{eqmodecomp} and (\ref{ec16}). 
For each of the set of differential coupled equations and their initial conditions, we have used 
 a fourth-order Runge-Kutta-Merson numerical scheme between $t=0$ and a maximum time $t_{\rm max}> 0$.
 In all cases, the perturbation is  turned on
for times $0<t<t_F$, with $t_F<t_{\rm max}$, where the system returns to a static configuration. 
For times $t<0$ and $t > t_F$, the cavity is a static one and we know  that the unperturbed solution can be written as 
in Eqs. (\ref{eps0}) and (\ref{epst}).

In order to compute the total number of particles created in a mode $n$, we follow the  procedure
of Ref. \cite{pre}. For $t \geq t_F$ the solution is of the form given  in Eq.(\ref{epst}).
We can therefore multiply both terms of the equation by $\exp(-i k_n t)$ and take the mean value in $t_F <t<t_{\rm max}$. 
In this way, we are able to numerically evaluate $|\beta_{nm}|^2$ and, also the particle number in mode 
$n$ as a function of time as $N_n(t_F) = \sum_m|\beta_{nm}(t_F)|^2$. 

The spectral modes $k_n$ are given in units of $1/d$ ($k_n d$ is dimensionless) and consequently time is measured in units of $d$. All figures are referred to dimensionless quantities.

\subsection {Equal driving frequencies $\Omega_R = \Omega_L = \Omega$}

\begin{figure}[h!]
\begin{center}
\includegraphics[width=8cm]{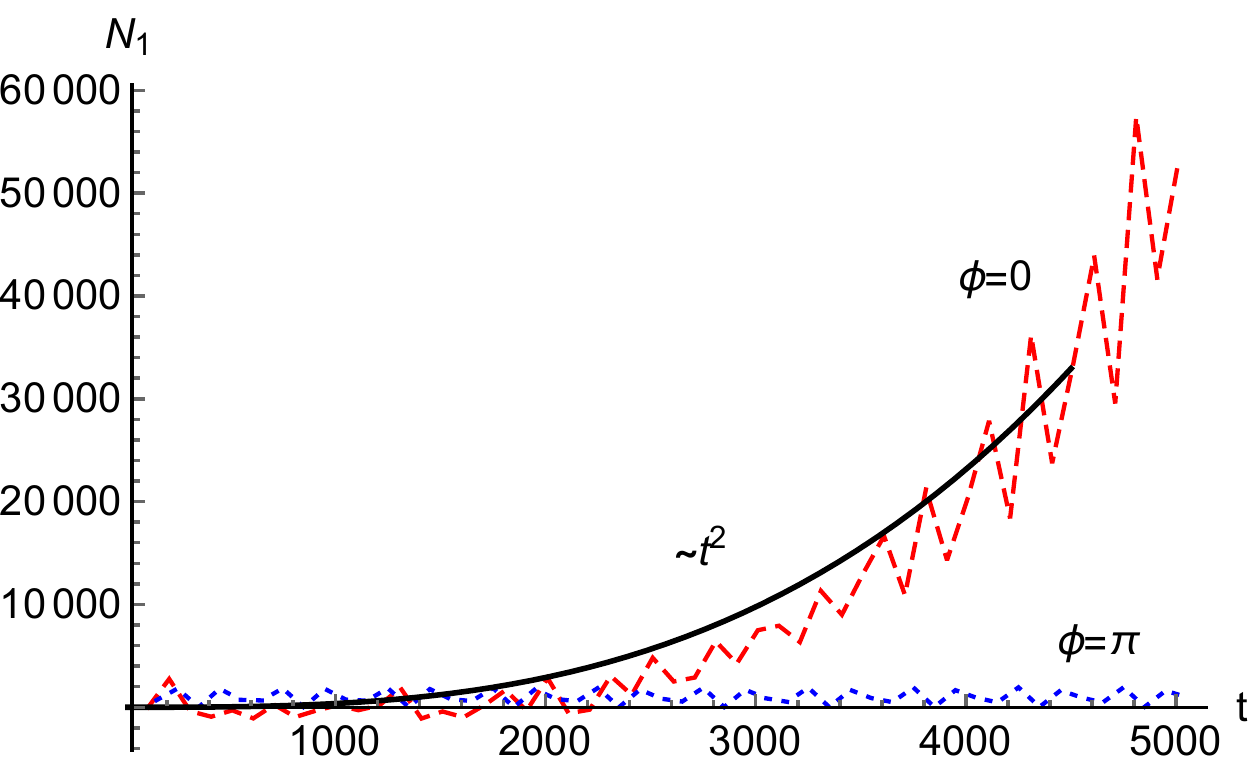}
\caption{(color online). Number of particle created in mode $n = 1$, $N_1$, for $b_0=500$ and relative phase $\phi =0$ (red dashed line). With dashed blue line, we show there is no particle creation for the translational mode with $\phi  = \pi$. Parameters used: $\Omega= 2k_1$, $\epsilon=0.01$, and $\chi_0 = 0.05$. Fit (solid black line) with the dimensionless time $t^2$.}
\label{caso500phi0}
\end{center}
\end{figure}

We begin by choosing big values for $b_{0L}$ and $b_{0R}$, for example $b_{0L}=b_{0R}= b_0 = 500$. In this case,  the particle creation is expected to behave quadratically with time \cite{param1}. In Fig.\ref{caso500phi0}, we can see the number of particles created in mode $n = 1$ for this situation when $\phi = 0$ and $\phi =\pi$ (we have set $\phi_R \equiv \phi$ the total relative phase). As expected for an equidistant spectra, the creation of particles grows quadratically with the time of excitation for breathing modes. On the contrary, for $\phi =\pi$, there is no photon creation (translational modes).

If we consider both values of $b_{0L}$ and $b_{0R}$ to be small and alike, we will be looking at  the non-equidistant region of the 
non-perturbed cavity spectrum (Fig.\ref{fig2}), \mbox{similar} to the situation described in Ref.\cite{suecos17}.

\begin{figure}[h!]
\begin{center}
\includegraphics[width=8cm]{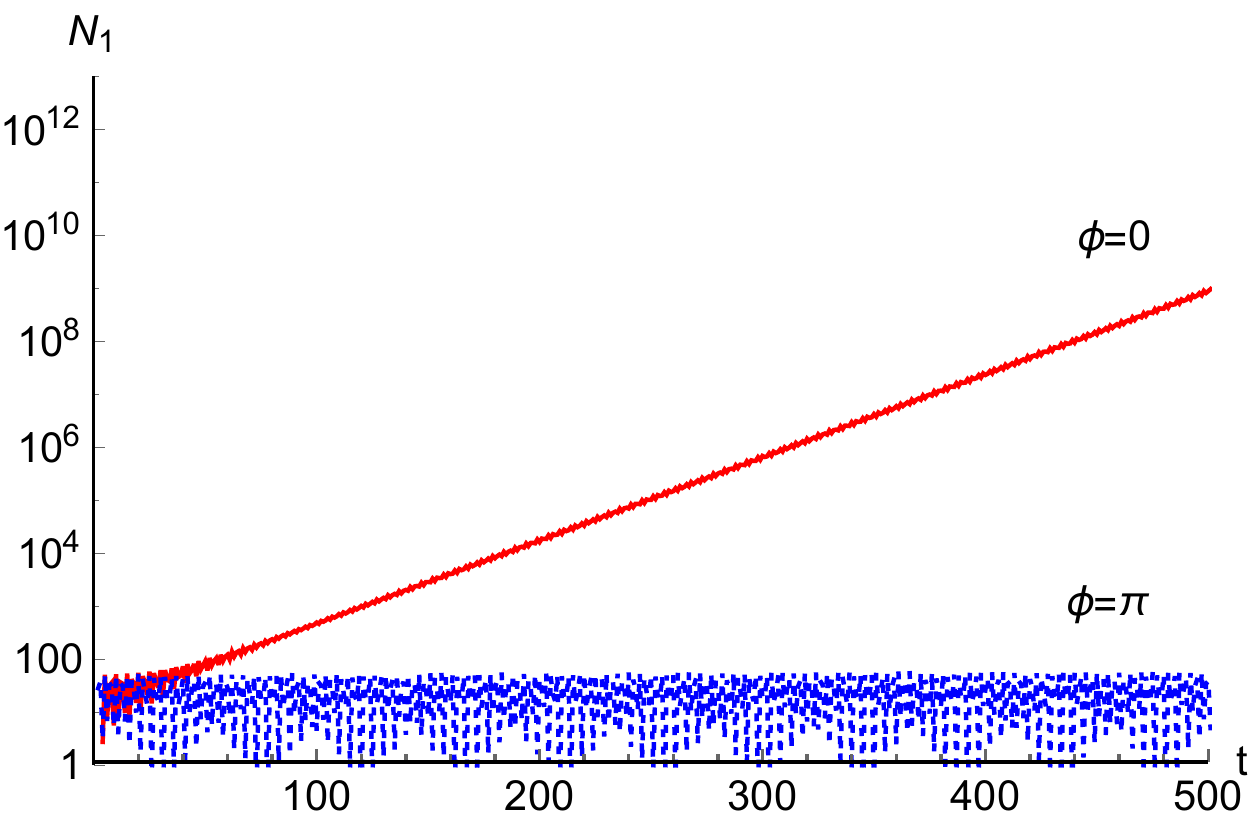}
\end{center}
\caption{(color online). Number of particle created in mode $n= 1$, $N_1$, for small and equal values of $b_0=1$, 
for $\phi =0$ and $\phi =\pi$. Parameters used: $\Omega= 2 k_1$, $\epsilon=0.01$, $\chi_0=0.05$.}
\label{caso1phi}
\end{figure}

We can hence get an insight of the photon creation inside the cavity for a non-equidistant spectrum. In Fig.\ref{caso1phi}, we show the number of particles created in field mode 1 ($N_1$) for an external perturbation $\Omega_R=\Omega_L= 2 k_1$. We can again note that there is no particle creation for translational modes. In the case of the breathing modes, the particle creation is exponential in time,  as expected. 

\begin{figure}[h!]
\begin{center}
\includegraphics[width=8cm]{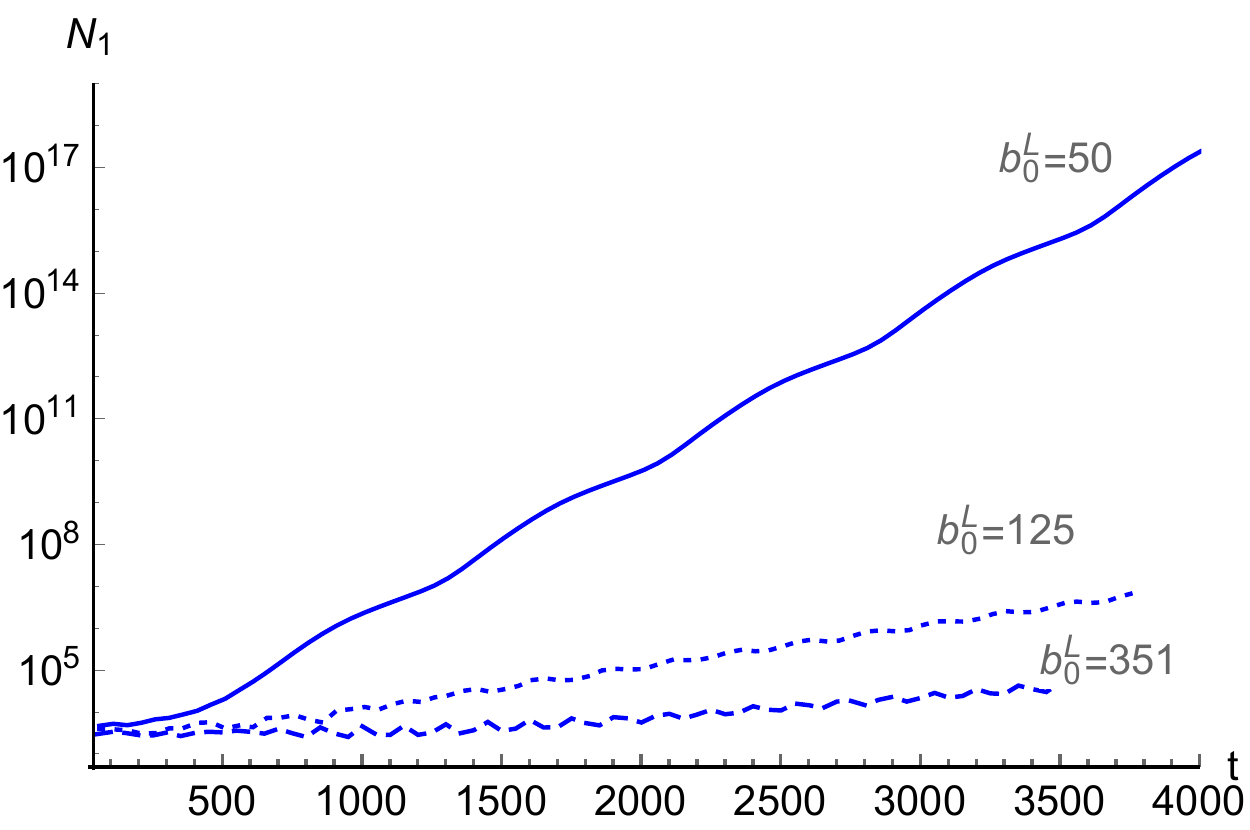}
\end{center}
\caption{(color online). Number of particles created in mode $n=1$, $N_1$, for different values of $b_{0L}$ and big value of $b_{0R}$. There is particle 
creation at short times, even for small values of $b_{0L}$. Parameters used: $\Omega= 2 k_1$, $\epsilon=0.01$, $b_{0R}=281$, $\phi=0$, 
and $\chi_0= 0.05$.}
\label{caso281phi0}
\end{figure}

In all intermediate regions of the cavity spectrum, the behavior will be as for a non-equidistant spectrum with different rates of particle 
creation as defined by the value of $b_0$.
In Fig.\ref{caso281phi0}, we show for example, different values of the number of particles created in field mode 1 ($N_1$), by setting different values of $b_{0L}$ and leaving fixed $b_{0R}=281$.
Therein, it is easy to see that the particle rate is bigger for lower values of $b_{0L}$ (the more non-equidistant region of the mode spectrum).

\begin{figure}[h!]
\begin{center}
\includegraphics[width=8cm]{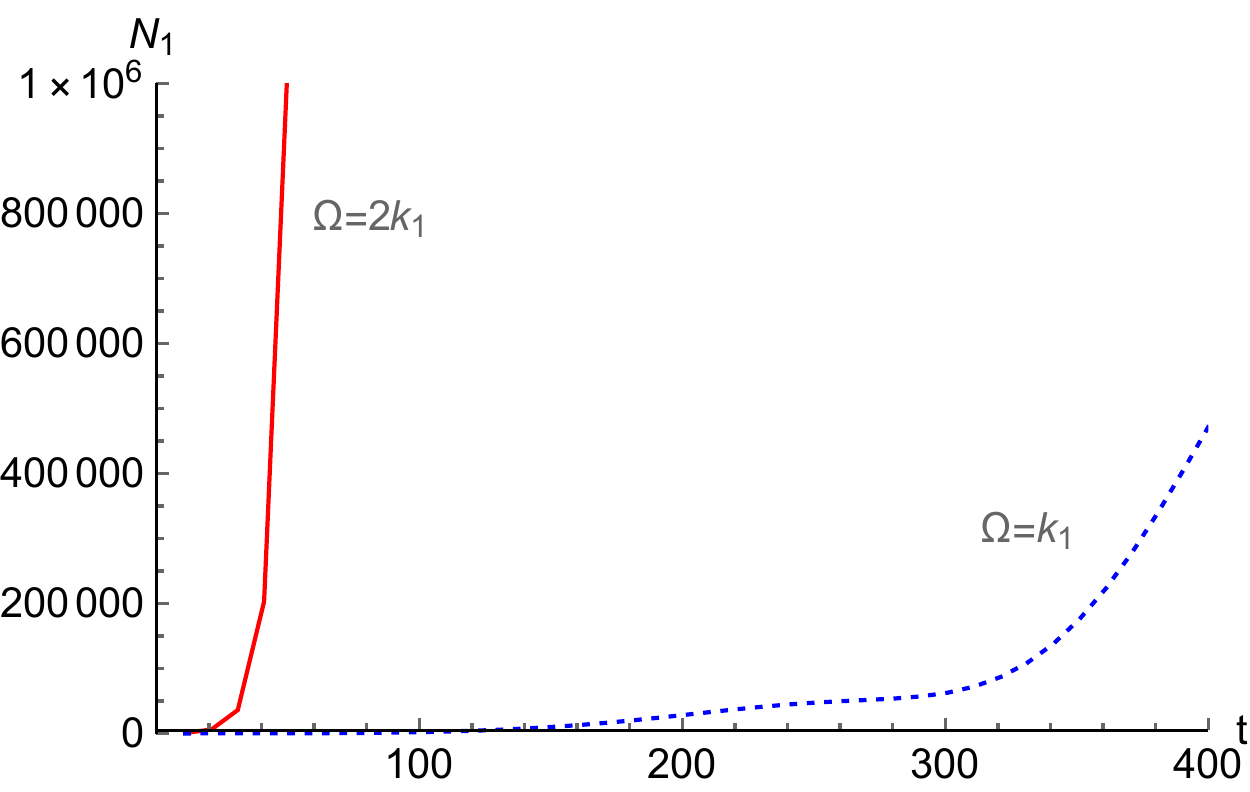}
\end{center}
\caption{(color online). Number of particle created in mode $n= 1$, $N_1$, for small and equal values of $b_0=1$, 
for different external frequencies: $\Omega= k_1$ and $\Omega= 2 k_1$. Parameters used: $\epsilon=0.05$, $\chi_0=0.05$.}
\label{comparacionresonancias}
\end{figure}

We shall now consider the number of particles \mbox{created} in mode 1 ($N_1$) for $\Omega_R=\Omega_L$ when exciting by $\Omega=2 k_1$ and compare it to the case when the external pumping frequency is $\Omega= k_1$ as shown in Fig.\ref{comparacionresonancias}. In this case, we are setting small values of $b_0=1$, $\epsilon=0.05$ and $\chi_0=0.05$. We can note that particle creation begins for times $t \sim 1/\epsilon \sim 20$ when the external pumping is $\Omega= 2 k_1$ while
the same occurs for times $ t \sim 1/\epsilon^2 \sim 400$ when $\Omega=k_1$. It is interesting to remark that the exponential 
growth  for  $\Omega=  k_1$ cannot be obtained analytically using the leading order of the MSA.

\begin{figure}[h!]
\begin{center}
\includegraphics[width=8cm]{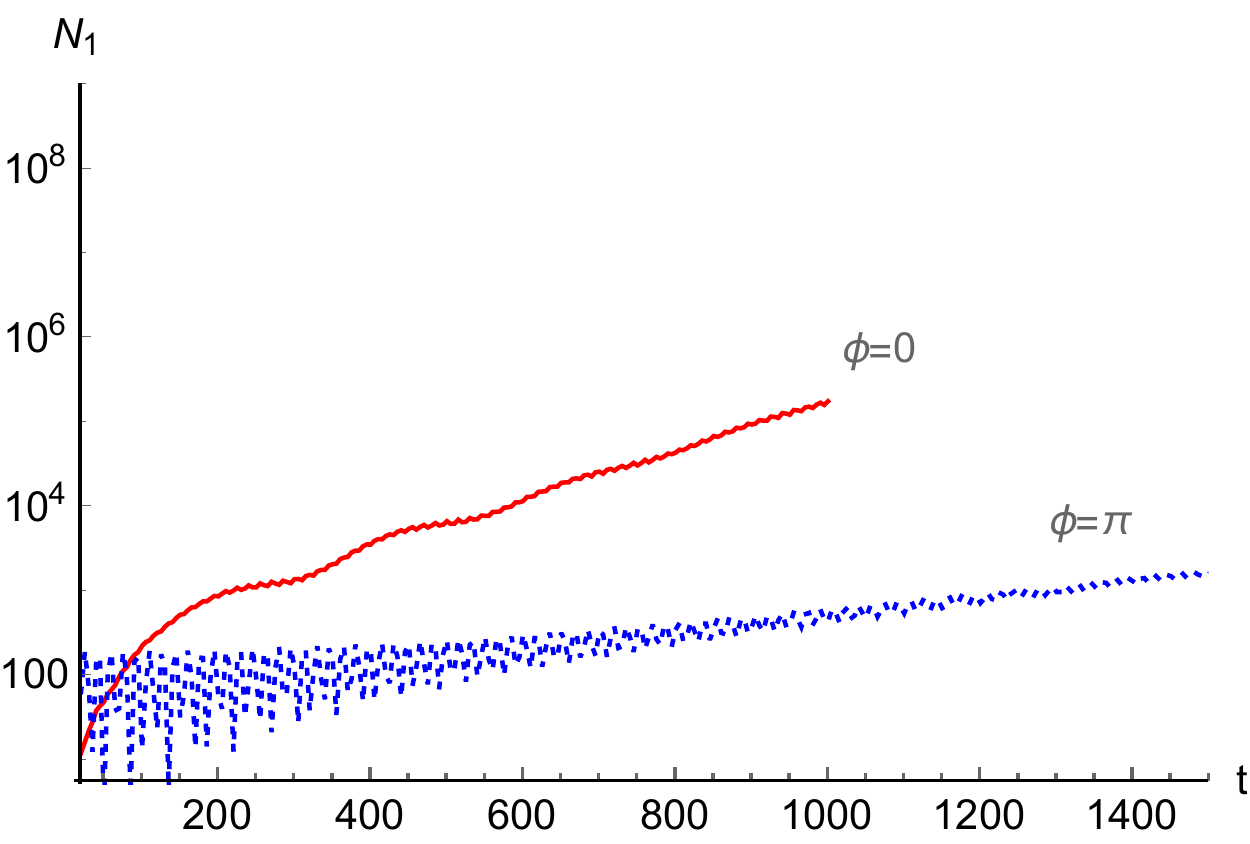}
\caption{(color online). Number of particles created in field mode 1 ($N_1$), for external frequencies $\Omega_R = \Omega_L= 2 k_1$ setting different values of $f_0^R=0.78$ and $f_0^L=0.46$ and $V_0^L=1.41$ and $V_0^R=5.59$, for $\phi=0$ (red solid line) and $\phi=\pi$ (blue dotted line). Parameters used: $\epsilon=0.01$ and $\chi_0=0.05$.}
\label{figura14}
\end{center}
\end{figure}

Finally,  we analyze the particle creation in mode field $n=1$ by setting different  initial values for $f_0^R$ and $f_0^L$. 
By choosing once more a non-equidistant region of the spectrum, we can set $V_0^L=1.41$, $V_0^R=5.59$, $f_0^L=0.46$ and $f_0^R=0.78$, yielding $b_{0L}=1$ 
and $b_{0R}=5$. The result is shown shown in Fig. \ref{figura14}. For these parameters, we  study the case $\Omega_R=\Omega_L= 2 k_1$ and see that there is particle creation in both cases, for $\phi=0$ and $\phi=\pi$ with a different rate accordingly analytical estimations. Indeed, from Eq.(\ref{onemode}) it is easy to see that there is no total destructive interference because, in this case, $\alpha_1^R \not= \alpha_1^L$.  On the contrary, in the other examples we presented along the paper, we have used $f_0^L= f_0^R=0.45 \pi$ and considered 
a small value for $\chi_0$. In those cases we have  $\alpha_1^R \simeq\alpha_1^L$,  and this is the reason why we have obtained  complete destructive interference.

\subsection {Different external frequencies: $\Omega_R \neq \Omega_L$}\label{difOmegas}

\begin{figure}[h!]
\begin{center}
\includegraphics[width=8cm]{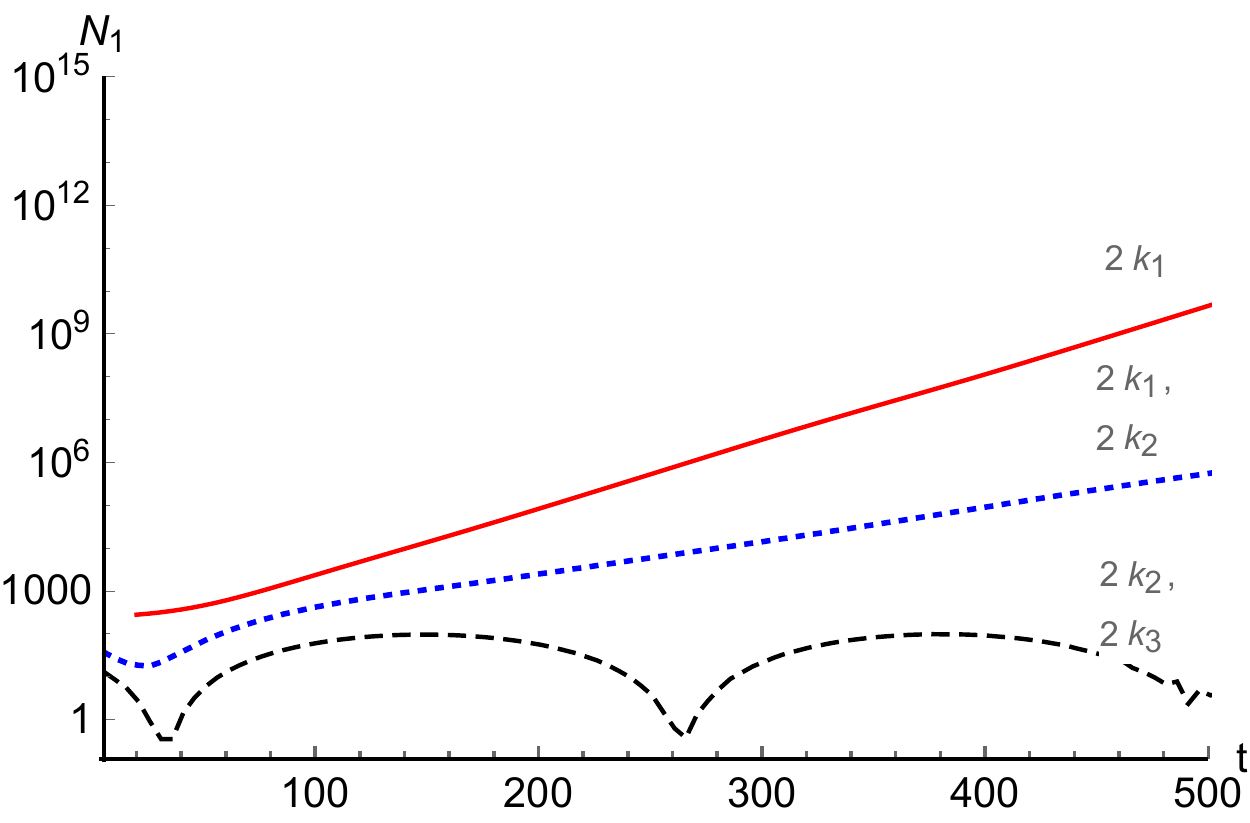}
\caption{(color online). Number of particle created in field mode $n= 1$, $N_1$, for different external frequencies.  $\Omega_R=\Omega_L= 2 k_1$ just for reference (red dotted line) and in the blue-dashed line the case $\Omega_R=2 k_1$  and  $\Omega_L= 2 k_2$ (overlapped with  $\Omega_R=2 k_2$ and $\Omega_L= 2 k_1$ ). Finally,  the black-dashed line corresponds to $\Omega_R=2 k_3$ and $\Omega_L= 2 k_2$. Parameters used: $b_0=1$, $\epsilon=0.01$ and $\chi_0=0.05$.}
\label{casofrecdistmodo1}
\end{center}
\end{figure}

In this Section we shall study the photon generation when the pumping frequencies are different, say $\Omega_R \neq \Omega_L$.
In Fig.\ref{casofrecdistmodo1} we show the number of created particles in field mode $n=1$, i.e. $N_1$, when the external frequencies are different
and given by 
 $\Omega_R=2 k_1$  and  $\Omega_L= 2 k_2$. In that figure we present the results overlapped with  $\Omega_R=2 k_2$ and $\Omega_L= 2k_1$, and added
$\Omega_R=\Omega_L= 2 k_1$ just for reference (red dotted line). In addition we show the particle creation in mode field 1 for  $\Omega_R=2 k_3$ and $\Omega_L= 2 k_2$ (black-dashed line).  We note that the field mode is excited only when is parametrically excited at least by one pumping frequency or one SQUID (indeed, black-dashed line in Fig.\ref{casofrecdistmodo1} shows no particle creation in the mode 1 as the cavity is 
excited with $\Omega_R=2 k_3$ and $\Omega_L= 2 k_2$).

\begin{figure}[h!]
\begin{center}
\includegraphics[width=8cm]{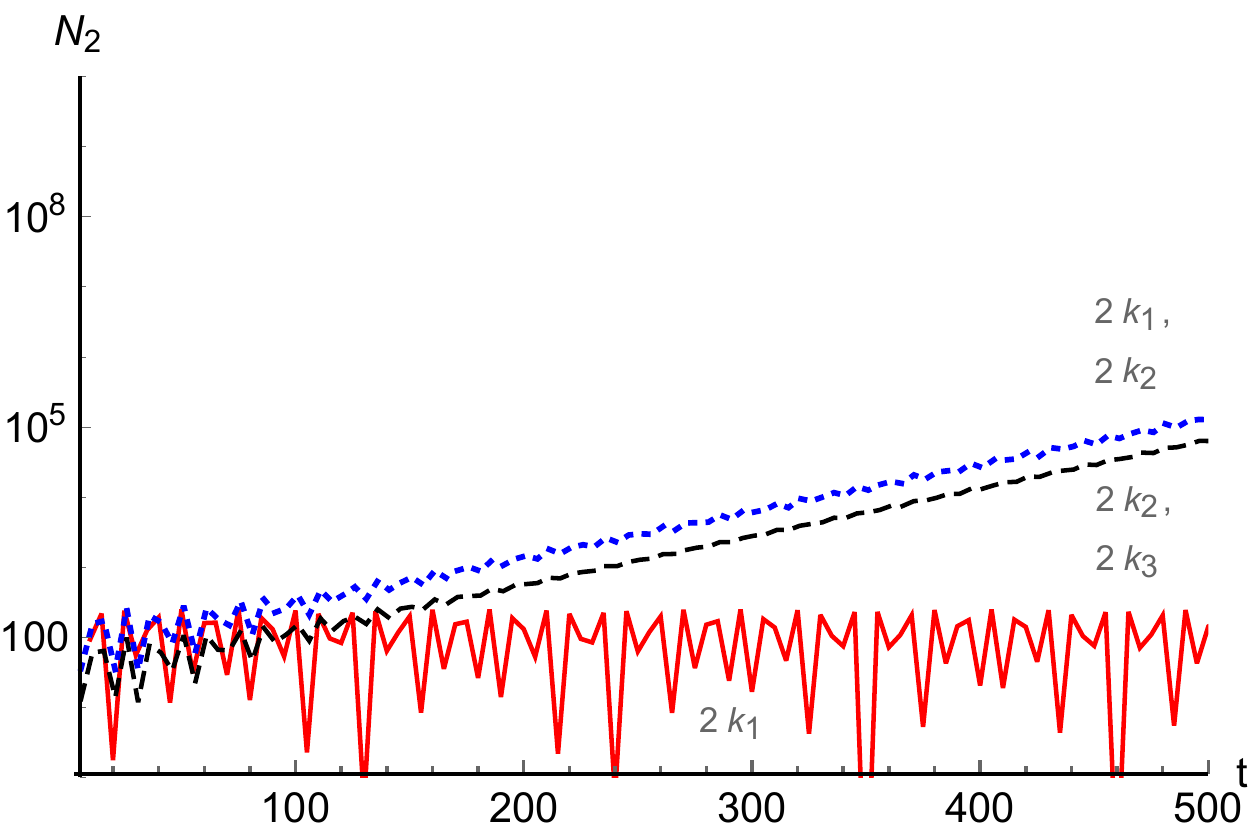}
\caption{(color online). Number of particle created in field mode $n= 2$, $N_2$, for different external frequencies. Dashed blue-line is $\Omega_R=2 k_1$ and  $\Omega_L= 2 k_2$, overlapped with $\Omega_R=2 k_2$ and $\Omega_L= 2 k_1$. The dashed-black line is for  $\Omega_R=2 k_3$ y $\Omega_L= 2 k_2$; while red solid-line corresponds to the case $\Omega_R=\Omega_L= 2 k_1$. Parameters used: $b_0=1$, $\epsilon=0.01$ and $\chi_0=0.05$.}
\label{casofrecdistmodo2y3}
\end{center}
\end{figure}

We see similar behaviours for the particle creation of field mode $2$ ($N_2$) in Fig.\ref{casofrecdistmodo2y3},  when 
the system is excited by different combinations of external frequencies. We see that in this case, there is no photon generation in modes $n =2, 3$ when $\Omega_R=\Omega_L= 2 k_1$ as expected.

\begin{figure}[h!]
\begin{center}
\includegraphics[width=8cm]{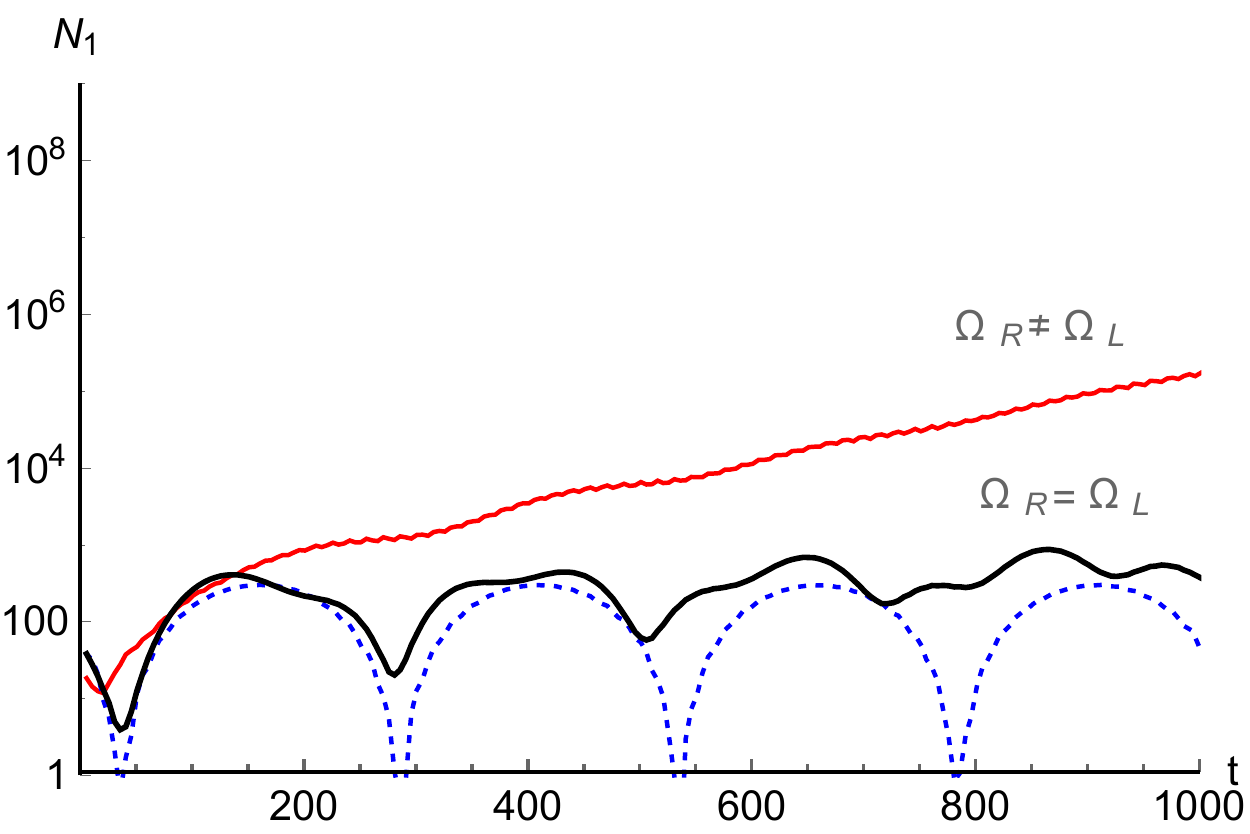}
\caption{(color online). Number of particles created in field mode $N_1$, for external frequencies are  $\Omega_R = k_2 + k_1$ and $\Omega_L = k_2 - k_1$. In the case of $\Omega_L = \Omega_R= k_2 - k_1$ there is no particle creation for $\phi=0$ and $\phi=\pi$. Parameters used: $\epsilon=0.01$ and $\chi_0=0.05$.}
\label{figura13}
\end{center}
\end{figure}

We can also present the number of created particles when the external frequencies satisfy that $\Omega_R=k_{n} \pm k_{m}$ and $\Omega_L=k_{n} \pm k_{m}$, whether they are in phase or not. In Fig. \ref{figura13}, we show different combinations of external excitations.  There is an exponential growth in $N_1$ when  $\Omega_R=k_{2} + k_{1}$ and $\Omega_L=k_{2} - k_{1}$, and 
no appreciable photon creation when $\Omega_R=\Omega_L=k_2 - k_1$, neither when $\phi=0$ nor $\phi=\pi$. These results 
has been anticipated by the analytic analysis of Section III.

\section {Detuning}

In this Section, we shall study the relevance of detuning in the  process of particle creation. We set parameters in the non-equidistant region of the 
spectrum and evaluate the number of created particles as function of the external driving frequencies. 

We have compared the case of an external perturbation $\Omega_L = \Omega_R = \Omega= 2 k_1$ and $\Omega_L = \Omega_R  = \Omega= k_1$ in Fig.\ref{comparacionresonancias}. Therein, we have seen that they differ in the timescale for which the particle creation begin and in the rate of particle creation.

\begin{figure}[h!]
\begin{center}
\includegraphics[width=8.5cm]{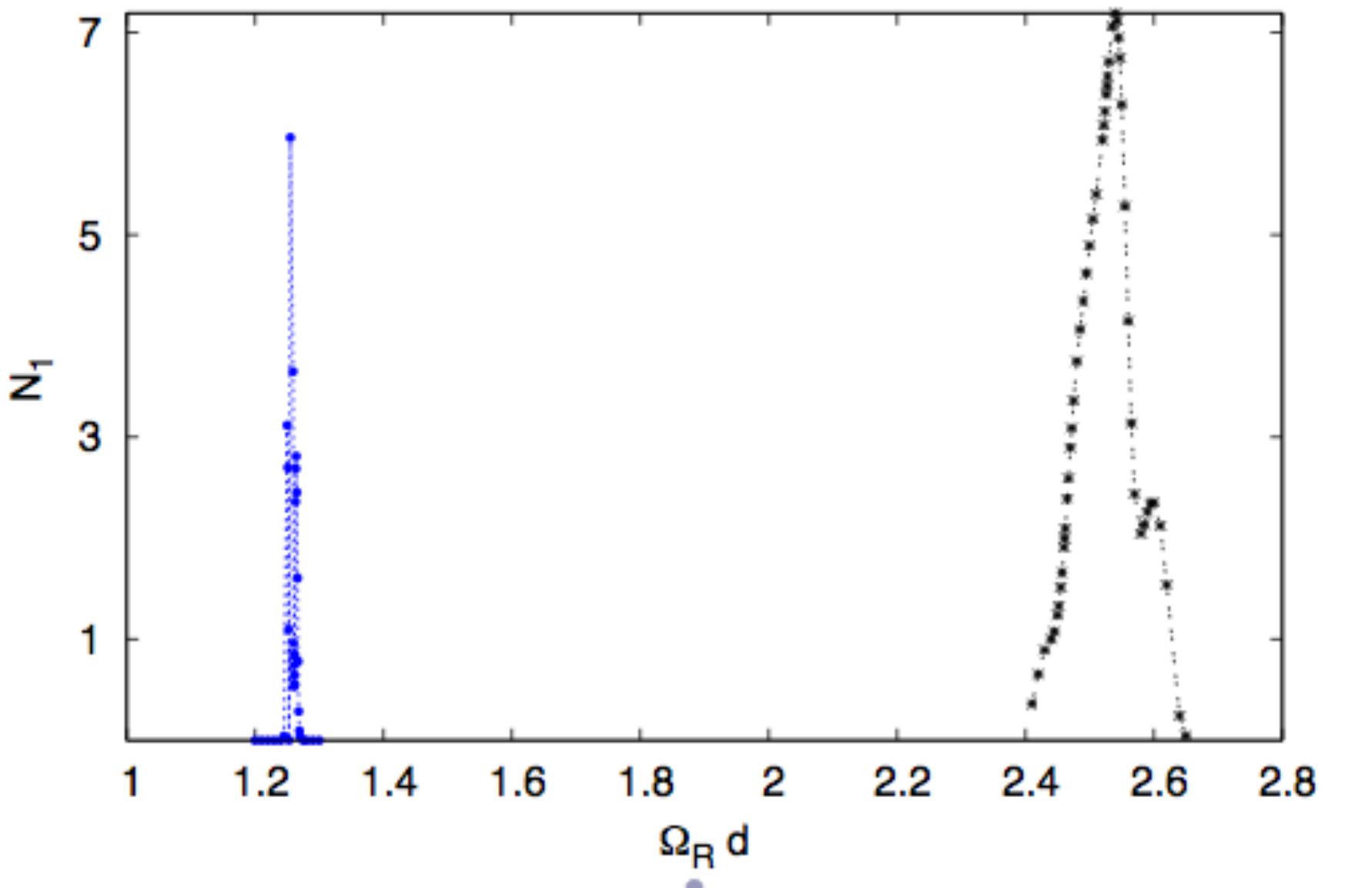}
\caption{(color online). Number of particle created in field mode 1 $N_1$ for $b_0=1$, for $\Omega=2 k_1$ and  $\Omega= k_1$, at different dimensionless times, for the case presented in Fig. \ref{comparacionresonancias}. In this figure, the number of created particles $N_1$ is plotted in units of $10^8$ for simplicity and clarity of the label.}
\label{picos}
\end{center}
\end{figure}

In Fig.\ref{picos}, we show the number of particles created for   $\Omega= 2 k_1$ and $\Omega= k_1$, with $\epsilon=0.05$ and $b_0=1$. When $b_{0R}=b_{0L} = 1$, the first eigenfrequency is $k_1=1.2611$ and the second one is $k_2= 3.3910$. As both perturbations determine different timescales, we compare the detuning process for the same number of particle created (obtained at different times in each case). This number of particles for field mode 1 is obtained for $t=80$ when $\Omega= 2 k_1$ and $t=410$ 
when $\Omega= k_1$ (see Fig.\ref{comparacionresonancias}). It is easy to note that the detuning is narrower in the case of a resonance of higher order. 

\begin{figure}
\begin{center}
                \includegraphics[width=8.5cm]{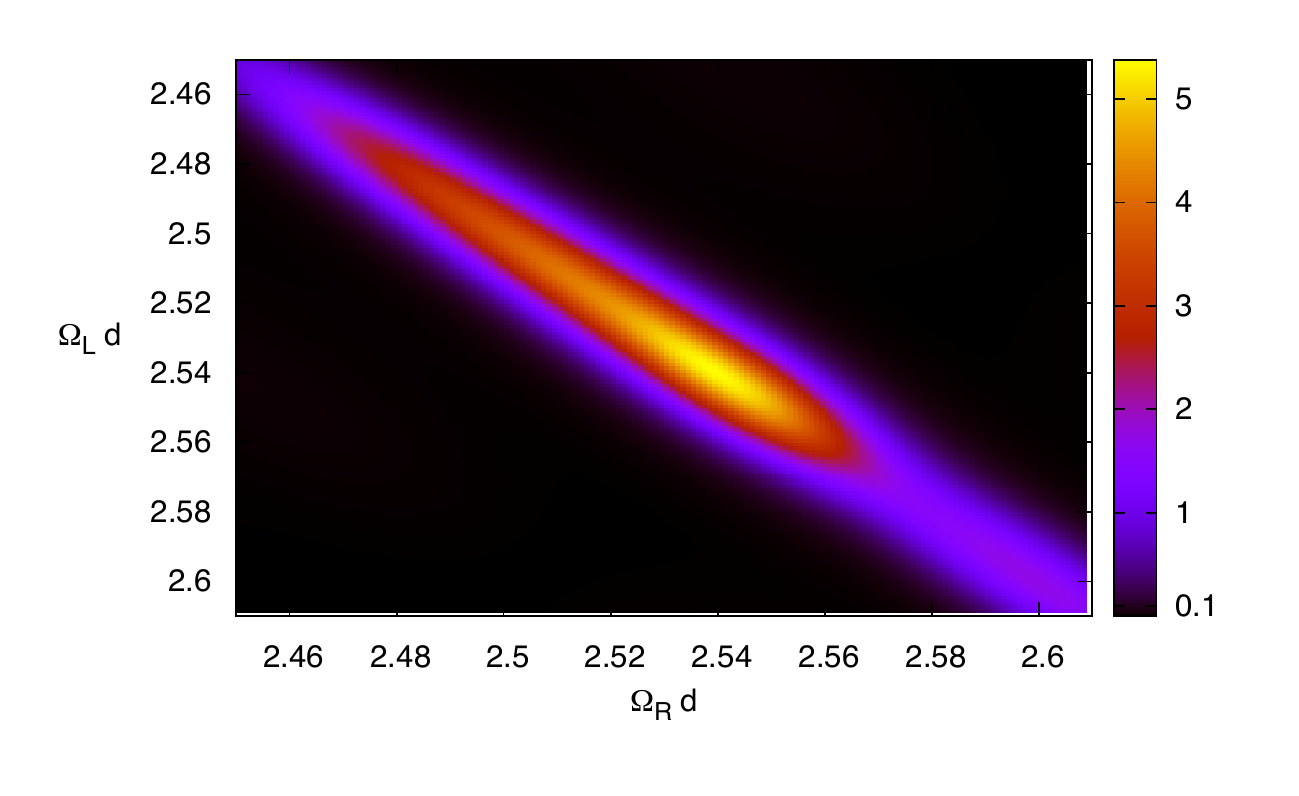}
               \caption{(color online). Number of particle created in field mode 1, $N_1$ for $b_0=1$, $\Omega= 2k_1$ at $t=80$. $N_1$ is plotted in units of $10^9$ for simplicity and clarity of the label.}
                \label{fig16}

      \end{center}
      \end{figure}

\begin{figure}
      \begin{center}
                \includegraphics[width=8.5cm]{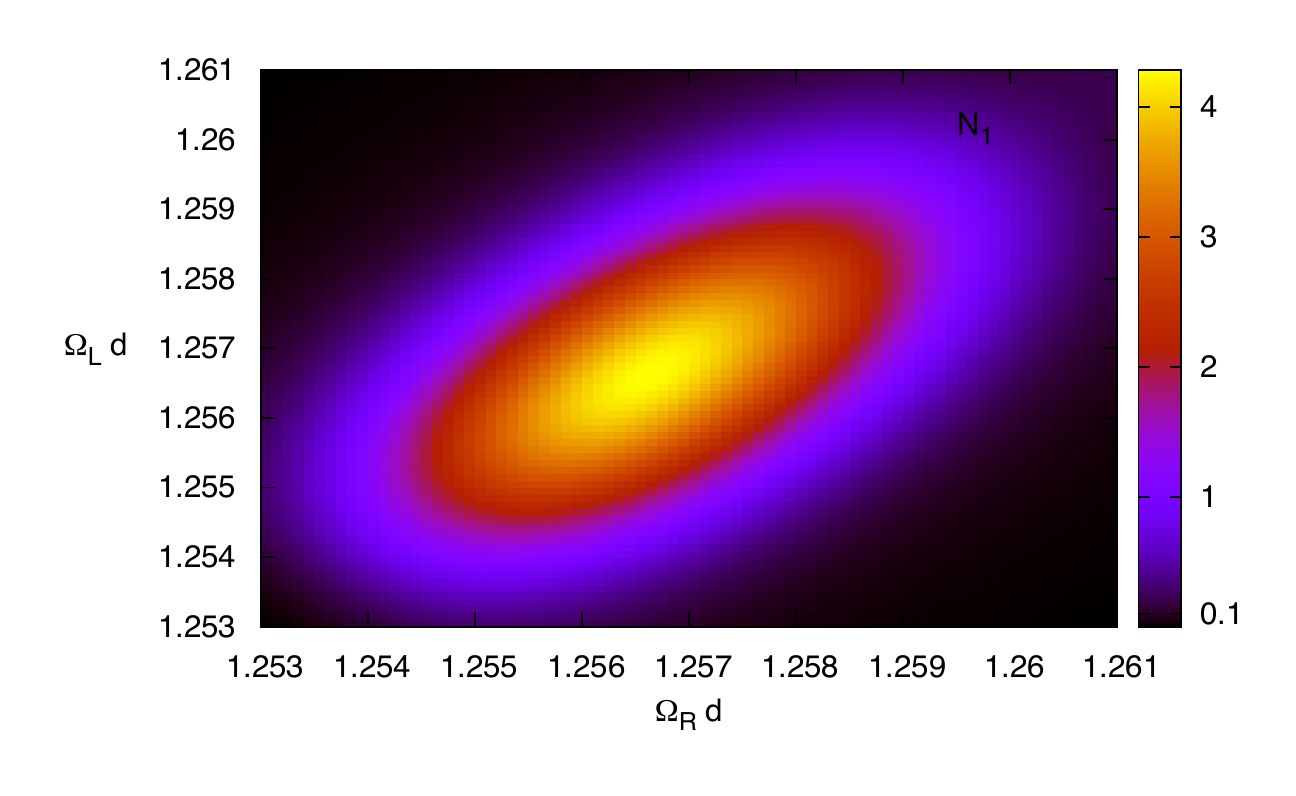}
                \caption{(color online). Number of particle created in field mode 1, $N_1$ for $b_0=1$, $\Omega= k_1$ at $t=410$. $N_1$ is plotted in units of $10^9$ for simplicity and clarity of the label.}
                \label{fig17}
\end{center}
\end{figure}

In  Figs.\ref{fig16} and \ref{fig17} we show the number of created particles as a function of both $\Omega_L$ and $\Omega_R$. In particular we show the detuning for the first eigenfrequency $n= 1$, when the external driving is $\Omega_L = \Omega_R = 2 k_1$ and $\Omega_L = \Omega_R = k_1$.

It is worthy noting that one can miss a resonance of higher order easier than the other, as the area covered is narrower.   

\begin{figure}[h!]
\begin{center}
\includegraphics[width=8.5cm]{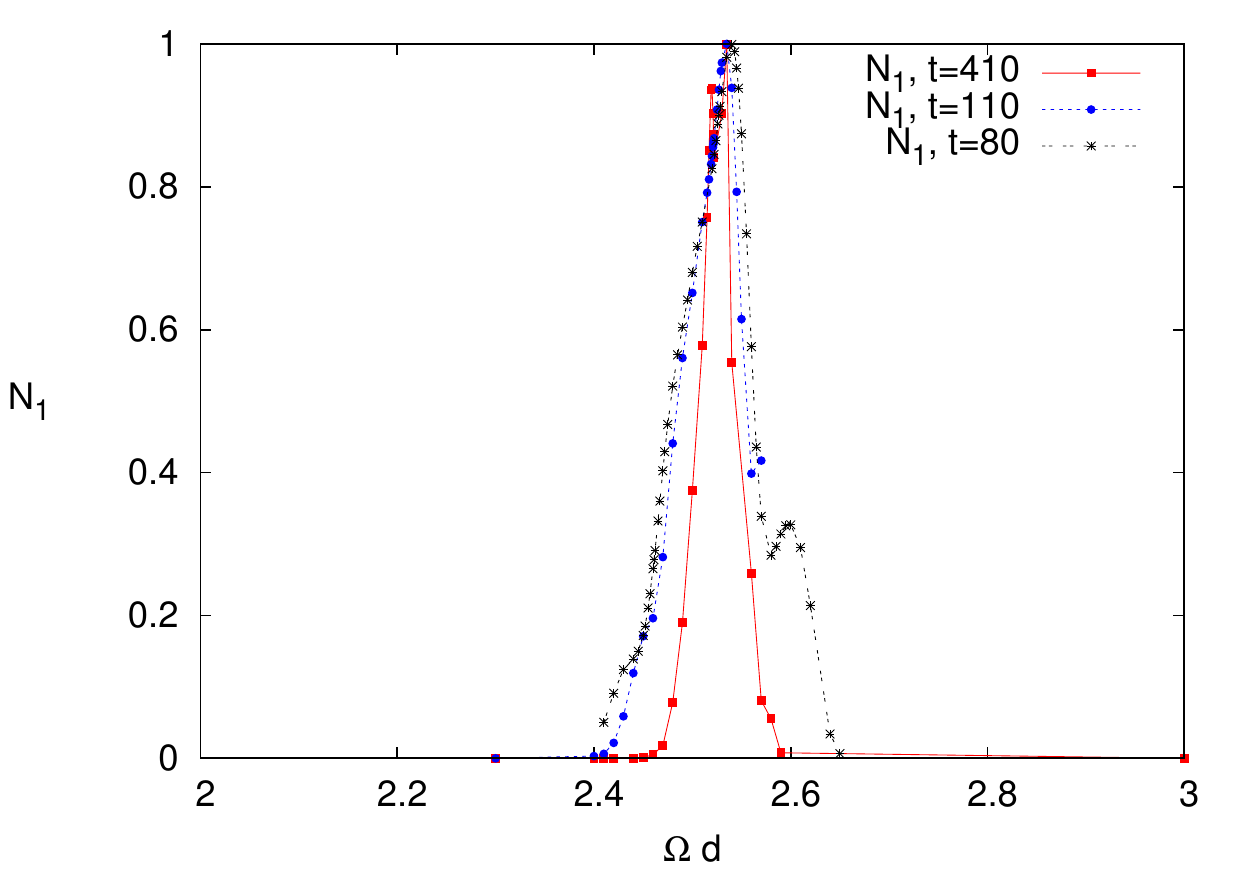}
\caption{(color online). Number of particle created in field mode 1 $N_1$ for $b_0=1$, for $\Omega=2 k_1$ at different times.}
\label{pico}
\end{center}
\end{figure}

Finally, we show that the structure of the peak response gets narrower as the time elapses. This is shown in Fig.\ref{pico} for $\Omega=2 k_1$ at different times: $t=80$, $t=210$ and $t=410$. The peaks are normalized in order to be compared at different times. 

\section{Conclusions}\label{sec:conc}

In this paper we presented an analytical and numerical  analysis of the particle creation in a tunable cavity ended with two SQUIDs, both subjected to external time dependent magnetic fields. We considered a situation in which the boundary conditions at both ends are periodic functions of time.

In order to get an analytical solution beyond naive perturbative calculations, in Section III we have studied the particle creation using MSA. 
We have shown that there is parametric resonance when the external frequencies are of the form $\Omega_{L,R}=2 k_n$ and/or $\Omega_{L,R}=k_n\pm k_m$ where
$k_n$ and $k_m$ are eigenfrequencies of the static cavity.  Under parametric resonance, the number of created particles grows exponentially, with a rate that depends not only on the amplitudes and frequencies of the external modulations, but also on the parameters of the static cavity. Moreover, the relative phase of the external modulation introduces interference effects in the rate of growth, in the sense that the number of created photons when two SQUDs are externally pumped is not the sum of the created particles by each individually pumped SQUID. 

From a numerical study of the spectrum of the tunable two SQUIDs cavity, in Section IV we found that 
with appropriate choices of the parameters of individual SQUIDs,  it is possible to generate equidistant or 
non-equidistant spectra. These different types of spectra result relevant to evaluate the particle creation rates. 

Section V was devoted to the numerical calculation of the particle creation rates. In addition to provide 
support to the analytic calculations of Section III, we investigated regimes which are non reachable with the lowest order MSA.  
For equal driving frequencies ($\Omega_R = \Omega_L$) and large values 
of parameter $b_0$, we showed that particle creation rate grows quadratically with the final time for breathing modes 
and that the particle creation is suppressed in the translational modes. On the other hand,  when setting the parameters 
of the static cavity in such a way that the spectrum becomes non-equidistant,  we found exponential rates for particle creation.
In this case we also found interference effects, and described situations in which the destructive interference is total (no exponential 
growth in the translational modes) and cases where it is partial (exponential growth with different rates both in breathing and translational modes). The 
amount of interference can be tuned  by adjusting the static magnetic fluxes on the SQUIDs.
We obtained similar results when the external frequencies are different, and found exponential growth of the number of created particles
not only for the usual case in which the frequencies are twice an eigenfrequency of the static cavity, but also when
they are given by the sum of wo modes $k_m+k_n$.

Finally, in Section VI we  investigated the dependence of the results with the tuning of the external frequencies,  an important
aspect for the experimental verification of these effects. Comparing the number of particles created in mode $n = 1$, 
for the cases $\Omega = 2 k_1$ and  $\Omega = k_1$ (when both driving frequencies are equal to each other), we have shown that 
 for the case $\Omega = k_1$, the tuning of the resonance is much more critical than in the case $\Omega = 2 k_1$, because its peak in frequencies is much narrower. This effect can in principle be analyzed analytically going beyond the leading order in the MSA, but the calculations
are rather cumbersome.

There are several interesting issues related to the present  work which deserve further analysis. The present case of a cavity ended 
by two SQUIDs not only introduce interference effects in the particle creation rate, as in the case of two moving mirrors \cite{Dalvit2,2walls}, 
but possible entanglement between pairs of photons generated from vacuum (see Refs. \cite{mikel} where it is shown that dynamical Casimir effect may generate multipartite quantum correlations).  In relation to  eventual variants of recent experiments \cite{Wilson,Paraoanu,suecos17}, 
a theoretical analysis including nonlinearities is also due.

\section*{Acknowledgements}
This work was supported by ANPCyT, CONICET, UBA and UNCuyo; Argentina. 


\begin{thebibliography}{bib}

\bibitem{Moore}G.T. Moore,  J. Math. Phys. {\bf 11},  2679 (1970).

\bibitem{Dodonov2010}
  V.~V.~Dodonov,
  Phys.\ Scripta {\bf 82} (2010) 038105.
  
  \bibitem{Dalvit2011}
  D.~A.~R.~Dalvit, P.~A.~Maia Neto and F.~D.~Mazzitelli,
  Lect.\ Notes Phys.\  {\bf 834} (2011) 419.
  
  \bibitem{Nation2012}
  P.~D.~Nation, J.~R.~Johansson, M.~P.~Blencowe and F.~Nori,
  Rev.\ Mod.\ Phys.\  {\bf 84} (2012) 1.
  
  
\bibitem{Yablo} E. Yablonovitch,  Phys. Rev. Lett.{\bf  62}, 1742 (1989).

\bibitem{Braggio}  A. Agnesi,  C. Braggio, G. Bressi, G. Carugno, F. Della Valle, G. Galeazzi, G. Messineo, 
F. Pirzio, G. Reali, and G. Ruoso, Journal of Physics: Conference Series {\bf 161}, 012028 (2009).  

\bibitem{Belgiorno} F.~Belgiorno, S.~L.~Cacciatori, G.~Ortenzi, V.~G.~Sala and D.~Faccio,
  Phys.\ Rev.\ Lett.\  {\bf 104}, 140403 (2010).
  
  \bibitem{analogue}
  R.~Schutzhold,
  arXiv:1110.6064 [quant-ph];
   N.~Westerberg, S.~Cacciatori, F.~Belgiorno, F.~Dalla Piazza and D.~Faccio,
  New J.\ Phys.\  {\bf 16} (2014) 075003.  
  
\bibitem{Nori}J.~R.~Johansson, G.~Johansson, C.~M.~Wilson and F.~Nori,
  Phys.\ Rev.\ Lett.\  {\bf 103} (2009) 147003.

\bibitem{Wilson} C.M. Wilson,  G. Johansson, A. Pourkabirian, M. Simoen, J. R. Johansson, T. Duty, F. Nori, and  
P. Delsing, Nature {\bf 479}, 376 (2011).
  
  \bibitem{Louko2015} J.~Doukas and J.~Louko,
  Phys.\ Rev.\ D {\bf 91}, 044010 (2015). 
 
 \bibitem{Paraoanu} P.~Lahteenmaki, G.~S.~Paraoanu, J.~Hassel and P.~J.~Hakonen,
  Proc.\ Nat.\ Acad.\ Sci.\  (2013). 
  
\bibitem{farina14} Andreson L. C. Rego, Hector O. Silva, Danilo T. Alves, and C. Farina, Phys. Rev. D{\bf 90}, 025003 (2014).
  \bibitem{1squid} F. C. Lombardo, F. D. Mazzitelli, A. Soba, and P. I. Villar
Phys. Rev. A {\bf 93}, 032501 (2016). 
  
  \bibitem{suecos17} I.M. Svensson, M. Pierre, M. Simoen, W. Wustmann, P. Krantz, A. Bengtsson, G. Johansson, J. Bylander, V. Shumeiko, P.  Delsing,  
  Journal of Physics: Conf. Series {\bf 969}, 012146 (2018).
  
  \bibitem{Dalvit2} D.~A.~R.~Dalvit and F.~D.~Mazzitelli,
  Phys.\ Rev.\ A {\bf 59}, 3049 (1999).

 \bibitem{2walls} P. I. Villar, A. Soba, and F. C. Lombardo, Phys. Rev. A {\bf 95}, 032115 (2017).
  
 \bibitem{Shumeiko}W. Wustmann and V. Shumeiko,
Phys. Rev. B {\bf 87}, 184501 (2013).  
  
  \bibitem{Fosco2013} C.~D.~Fosco, F.~C.~Lombardo and F.~D.~Mazzitelli,
  Phys.\ Rev.\ D {\bf 87}, 105008 (2013).

\bibitem{param1} V.V. Dodonov and A.B. Klimov, Phys.
Rev. A {\bf 53}, 2664 (1996).

\bibitem{param2}  A. Lambrecht,  M.T. Jaekel and S. Reynaud, Phys. Rev. Lett.{\bf  77},  615 !996).  
   
   \bibitem{error} In Ref.\cite{1squid} the number of created particles was evaluated just from the diagonal part of $\epsilon_{nm}$. The results 
   do not differ appreciably under resonance conditions. 
   
 \bibitem{bender} C.M.  Bender and S.A. Orszag, {\it Advanced Mathematical Methods for Scientists and Engineers}, McGraw-Hill, Inc., New York (1978).
      
   \bibitem{crocces}M.~Crocce, D.~A.~R.~Dalvit and F.~D.~Mazzitelli,
  Phys.\ Rev.\ A {\bf 64}, 013808 (2001); ibidem  Phys.\ Rev.\ A {\bf 66}, 033811 (2002).
  
   \bibitem{pre} P. I.Villar and A. Soba, Phys. Rev. E {\bf 96}, 013307 (2017).
    

\bibitem{mikel} S. Felicetti, M. Sanz, L. Lamata, G. Romero, G. Johansson, P. Delsing, and E. Solano, Phys. Rev. Lett. {\bf 113}, 093602 (2014); 
D. Z. Rossatto, S. Felicetti, H. Eneriz, E. Rico, M. Sanz, and E. Solano, Phys. Rev. {\bf B} 93, 094514 (2016). 

 \end{thebibliography}
\end{document}